\documentclass[useAMS,usenatbib,usegraphicx]{mn2e}

\def\apj{ApJ}
\def\apjs{ApJS}
\def\apjl{ApJL}
\def\mnras{MNRAS}
\def\aj{AJ}
\def\nat{Nature}

\def\physrep{Phys.Rep.}      

\newcommand{\pot}[2]{#1 \times 10^{#2}}
\newcommand{\runit}{~{\rm h}^{-1}~{\rm Mpc}}
\newcommand{\munit}{~{\rm h}^{-1}~{\rm M}_\odot}
\newcommand{\degree}{\ensuremath{^\circ}}
\newcommand{\vpec}{v_{\rm p}}

\usepackage{amsmath}
\usepackage{amssymb}
\usepackage{tabularx}
\usepackage{color}

\title[SZ effect in superclusters: CrB]{The Sunyaev-Zeldovich effect in
  superclusters of galaxies using gasdynamical simulations: the case of Corona
  Borealis}

\author[I.~Flores-Cacho et al.]{I.~Flores-Cacho$^{1,2}$,
J.A.~Rubi\~no-Mart\'in$^{1,2}$, G.~Luzzi$^{3}$, R.~Rebolo$^{1,2,4}$, M.~De
Petris$^{3}$, \newauthor G.~Yepes$^5$, L.~Lamagna$^{3}$, S.~De
Gregori$^{3}$, E.S.~Battistelli$^{3}$, R.~Coratella$^{3}$ \newauthor
and S.~Gottl\"ober$^{6}$\\ \\ $^1$Instituto de Astrof\'isica de
Canarias, 38200 La Laguna, Tenerife, Spain\\ $^2$Departamento de Astrof\'isica, Universidad de La Laguna, E-38205 La Laguna, Tenerife, Spain\\ $^3$Department of
Physics, University di Roma "La Sapienza", Piazzale Aldo Moro 2, 00185
Roma, Italy\\ $^4$Consejo Superior de Investigaciones Cient\'ificas,
Spain\\ $^5$Grupo de Astrof\'isica, Universidad Aut\'onoma de Madrid,
Madrid E-28049, Spain\\ $^6$Astrophysikalisches Institut Potsdam, An
der Sternwarte 16, 14482 Potsdam, Germany }

\begin{document}

\date{Received **insert**; Accepted **insert**}

\pagerange{\pageref{firstpage}--\pageref{lastpage}}
\pubyear{}

\maketitle

\label{firstpage}

\begin{abstract}
We study the thermal and kinetic Sunyaev--Zel'dovich (SZ) effect
associated with superclusters of galaxies using the MareNostrum Universe
SPH simulation.  In particular, we consider superclusters with
characteristics (total mass, overdensity and number density of cluster
members) similar to those of the Corona Borealis Supercluster (CrB-SC).
This paper is motivated by the detection at 33~GHz of a strong temperature decrement
in the cosmic microwave background towards the core of this supercluster
(G\'enova-Santos et al. 2005, 2008).
Multifrequency observations with VSA and MITO suggest the existence of a thermal
SZ effect component in the spectrum of this \emph{cold spot}, with a
Comptonization parameter value of $y=7.8^{+4.4}_{-5.3}\times10^{-6}$
(Battistelli et al. 2006), which would account for roughly 25 per cent
  of the total observed decrement.

From the SPH simulation, we identify nine $(50$~$\runit)^3$ regions
  containing superclusters similar to CrB-SC, obtain the associated SZ maps and
calculate the probability of finding such SZ signals arising from hot gas within
the supercluster.
Our results show that the Warm--Hot Intergalactic Medium (WHIM) lying in the
intercluster regions within the supercluster produces a thermal SZ
effect much smaller than the observed value by MITO/VSA.

Neither can summing the contribution of small clusters and galaxy groups ($M <
\pot{5}{13}$~$\munit$) in the region explain the amplitude
of the SZ signal. Our synthetic maps show peak $y$-values significantly 
below the observations. Less than 0.3\% are compatible at the lower end 
of the 1-sigma level, even when considering privileged orientations in 
which the filamentary structures are aligned along the line of sight.
When we take into account the actual posterior distribution
from the observations, the probability that WHIM can cause a thermal
SZ signal like the one observed in the CrB-SC is $<1\%$, rising up to
a $3.2\%$ when the contribution of small clusters and galaxy groups is
included.
If the simulations provide a suitable description of the gas physics,
then we must conclude that the thermal SZ component of the CrB spot
most probably arises from an unknown galaxy cluster along the line of 
sight.
On the other hand, the simulations also show that the kinetic SZ signal
  associated with the supercluster cannot provide an explanation for the
  remaining 75\% of the observed cold spot in CrB.
\end{abstract}

\section{Introduction}

A detailed account of the baryons present in all known components of the local
Universe gives a value for the baryon density parameter of $\Omega_{\rm
B}=(0.010\pm0.003)h^{-2}$ \cite[see e.g.][]{FHP98}). Primordial nucleosynthesis
\citep{BNT01}, observations of the Lyman-$\alpha$ forest \citep{Rauchetal97} and
measurements of the angular power spectrum of the Cosmic Microwave Background
\cite[CMB,][]{Rebolo2004,Spergel2006} lead to values  two times
higher for this parameter.
It seems that there must exist a baryonic component in the local Universe where
half of the baryons would remain hidden. 
Theoretical models \citep[e.g.][]{CenOstriker99} based on gasdynamical
simulations suggest that these missing baryons could be accounted for in a
diffuse gas phase with temperatures $10^5 < T < 10^7$K and moderate
overdensities ($\delta \la 10-100 $), known as the \textquoteleft warm/hot
intergalactic medium' (WHIM). According to these simulations, the WHIM would be
located in filaments connecting clusters of galaxies and in large-scale
sheet-like structures.
The existence of the WHIM component is a fairly robust conclusion against
changes in the gas physics included in the simulations, as feedback processes
\cite[see e.g.][]{CenOstriker06}. It has been shown that even when considering
nonequilibrium for major metal species, UV photoionization and galactic
superwinds, the fraction of baryons in the form of WHIM is still of order 

%
There is observational evidence of this WHIM component, using
absorption lines in the spectra of background sources, either in
ultraviolet wavelengths for the lower temperature WHIM
\citep[e.g.][]{Savage1998, Nicastro2003}, or in the X-ray range (first
claimed detection by \citet{Nicastro2005} towards Mkn 421, although it
is still under discussion; recently detected by \citet{Buote2009}
towards the Sculptor Wall). Apart from this possibility, the WHIM could
also be detected via inverse Compton scattering of CMB photons, the
so-called Sunyaev-Zel'dovich (SZ) effect \citep{SZ72}. In this work we
focus on this physical process and study both the thermal (tSZ)
and kinetic (kSZ) signals associated with this gas phase.
The SZ effect was first proposed for clusters of galaxies and has
since been proven to be an extremely useful tool for studying cluster
physics. When combining SZ measurements with X-ray and/or optical
measurements it is possible to obtain an independent determination of
the cosmological parameters \citep{Carlstrom00}. Recently, it has also
been proven that SZ surveys are effective means of finding new galaxy
clusters in blind surveys \citep{Staniszewski08}.

Although the SZ effect has only been observed robustly in the direction of
clusters of galaxies \cite[for a review see][]{Birkinshaw99},
the aforementioned simulations by \citet{CenOstriker99} and other theoretical
models \citep{Persicetal88, Persicetal90, Boughn99} suggest that SZ surveys in
superclusters could lead to a detection of the WHIM.
The basic idea is that since the SZ effect is proportional to the line-of-sight
integral of the electron pressure, filamentary structures of superclusters
extending over several tens of megaparsecs could overcome the expected low
baryon overdensities and produce a measurable signal with present-day observing
facilities.
Following these ideas, \citet{Atrio06} theoretically computed the
expected SZ anisotropies associated with this WHIM phase and found its
dependence on gas and cosmological parameters. Moreover,
and based on numerical simulations, \citet{Dolagetal05, Hansenetal05,
  HMetal06} and \citet{Roncarellietal07} have provided coinciding
predictions of the contribution of diffuse (non-collapsed) gas to the
thermal SZ power spectrum. In particular, \citet{Hallman07} concluded
that the contribution from unbound gas could be as high as 12\% that
of galaxy clusters.

%
Observations of the Corona Borealis (CrB) supercluster carried out
with the extended configuration (11~arcmin resolution) \citep{Watson_et_al_2003}
of the Very Small Array (VSA) at 33~GHz \citep{Ricardo2005} showed the existence
of two strong and resolved negative features in a region where there
are no known clusters of galaxies.
Our study focuses on the strongest spot, the so-called ``H-spot'', which has a
temperature decrement of $\Delta T = -230 \pm 23 ~\mu$K.
This region is in fact the most negative feature found in any of the
  maps produced by the VSA, which in total covered a sky area of $\sim
  100$~deg$^2$. Moreover, a detailed Gaussianity study in the region
\citep{Rubinoetal06} finds a clear deviation (99.82\%) at angular scales of
multipole $\ell \sim 500$.
\citet{Ricardo08} have recently presented new VSA observations with an equivalent
angular resolution of 6~arcmin that confirm the existence of the ``H-spot''.

The origin of the CrB spot is still unclear. Among other options, a
  possible explanation could be an SZ effect.
Observations performed with the MITO (Millimetre \& Infrared Testa Grigia
Observatory) telescope \citep{dePetris_99} at 143, 214 and 272~GHz and an
angular resolution of 16~arcmin showed that roughly 25\% of the total
  signal has a thermal SZ spectral behaviour, yielding a Comptonization
  parameter of $y=7.8^{+4.4}_{-5.3}\times10^{-6}$ at 68\% C.L. \citep{Elia2006,
    Elia2007}.
Summarizing, we need to find an explanation for this tSZ component, but
  we also need to understand the origin of the remaining $\sim 75$\% of the cold
  spot: even when removing the detected tSZ signal, the CrB-H decrement is a
  $3.9\sigma$ deviation with respect to Gaussianity.
%

%
In this paper, we use numerical simulations to explore the possibility that the
signal observed in the CrB-H spot  originates by diffuse gas (WHIM) within the
supercluster. Basically, we want to test whether or not the
MNU simulations are able to explain the detected tSZ signal by MITO/VSA in terms 
of the WHIM. Additionally, we test whether the same WHIM component can provide a relevant 
contribution to the total decrement in terms of its kinetic SZ effect.
N-body numerical gasdynamical simulations have proved a very useful tool
in establishing accurate theoretical predictions of the SZ signals
induced either by galaxy clusters or the WHIM component \citep[see
  e.g.][]{DaSilvaetal00, Springeletal01, Dolagetal05, Hansenetal05,
  HMetal06, Roncarellietal07, Hallman07, Hallman09}.
However, these studies have mainly focused on the statistical properties of
these signals (e.g. predictions for the SZ angular power spectrum, or the
one-point probability distribution function of the full sky maps as a whole).
In this study, we restrict our analyses to superclusters and we study the
probability of finding spots produced by SZ signals such as those observed in CrB.
In addition, we also provide predictions for the observability of those
individual features in the maps produced by the the next generation of
instruments, such as the Planck satellite \citep{planck}.

In \S\ref{section2} we briefly describe the main characteristics of the
simulations used; in \S\ref{section3} a complete description of the methodology
applied in the study is given, and in \S\ref{section4} we present a
  general statistic description of the maps. In \S\ref{section5} we present our
  main analysis for the particular case of the CrB-SC spot while in \S\ref{section6} 
  we provide an estimate of the observability of SZ signals with Planck.
  Finally, in \S\ref{section7} we present a discussion and our conclusions obtained 
  with this study.

\section{Simulations}
\label{section2}

For the analysis reported in this paper we have used the
\emph{MareNostrum Universe} (hereafter MNU) SPH simulation, which is currently
the largest cosmological N-body+SPH simulation carried out so far. It
consists of $1024^3$ dark and $1024^3$ SPH particles in a cubic
computational volume of $500\runit$ on a
side.\footnote{http://astro.ft.uam.es/$\sim$marenostrum}  This
simulation was done at the MareNostrum supercomputer, from which it
took the name, located at the Barcelona Supercomputer Center, Centro
Nacional de Supercomputaci\'on
(BSC-CNS).\footnote{http://www.bsc.es} Initial conditions were set up
according to the so-called ``Concordance $\Lambda$-CDM Model'' with
parameters $\Omega_{\rm m}=0.3$, $\Omega_{\Lambda}=0.7$, $\Omega_{\rm
  b}=0.045$, $\sigma_{8}=0.9$ and $H_0 = 70$~km~s$^{-1}$~Mpc$^{-1}$
and a slope of $n=1$ for the initial power spectrum (see
\citet{GottloeberYepes07} and \citet{Yepes07} for more details of
this simulation).

The mass of each dark matter particle is $m_{\rm DM} = 8.239 \times
10^9 \munit$ and the mass of each gas particle is $m_{\rm gas}=1.454
\times 10^9 \munit$, evolved from redshift $z=40$ using the TREEPM+SPH
code GADGET-2 \citep{Springeletal01, Springel05}.
The spatial force resolution is set to an equivalent Plummer
gravitational softening of $15$~h$^{-1}$~kpc, and the SPH smoothing
length is set to the 40th neighbour to each particle.

We focused our work on the snapshot corresponding to redshift $z=0$,
for which complete catalogues of objects are available, as described
in the next subsection.
Finally, to investigate the dependence of the results on the mass and
spatial resolutions of the simulations, we have also considered a
re-simulation of the same snapshot using identical initial conditions
but with $2\times256^3$ particles, which are also equally distributed
between gas and dark matter, with $m_{\rm
  gas}=8.065\times10^{10}$~$\munit$ and $m_{\rm
  DM}=5.397\times10^{11}$~$\munit$. The same halo catalogues were available for this lower resolution
simulation.

\subsection{Halo identification algorithms}
\label{section2.1}

In order to find all structures and substructures within the
distribution of 2 billion particles and to determine their properties
we have used two different algorithms: a friends-of-friends analysis and
a spherical overdensity halo-finder. We have started with the parallel
version of the hierarchical friends-of-friends (FOF) algorithm
\citep{Klypin99}. We use a basic linking length of 0.17 of the mean
interparticle separation to extract the FOF objects at redshift $z=0$.
With this linking length we have identified more than 2 million
objects with more than 20 DM particles which closely follow a
Sheth-Tormen mass function \citep{GottloeberYepes06}. More than 4000
galaxy clusters with masses larger than $10^{14} \munit$ have been found. 
If one divides this linking length by $2^n$ ($n=1,3$) substructures 
and in particular the centres (density peaks) of the objects can be 
identified.

Spherical halos are identified using the AMIGA-Halo-Finder (AHF)
\citep{ahf}, an MPI parallelized modification of the
algorithm presented in \citet{Gilletal04}. The halo finder locates halos
(as well as subhalos) as peaks in an adaptively smoothed density
field. The local potential minima are computed for each peak and
within spherical volumes the set of particles that are gravitationally
bound to the peak is determined.  If the peak contains more than 20
particles, then its virial (or truncation) radius is calculated. The
virial radius $R_{\rm vir}$ is the point at which the density profile
drops below the virial overdensity $M(<R_{\rm vir})/(4\pi R_{\rm
vir}^3/3) = \Delta_{\rm vir} \rho_{\rm mean}$, where $\Delta_{\rm vir}$ depends
on the cosmology \citep{Kitayama96}; in our case, $\Delta_{\rm vir} = 330$. 
However, for subhalos this prescription is not
appropriate.  Subhalos exist in the dense environment of their host
halo, where the density exceeds the virial one.  In this case the
density profile shows a characteristic upturn. This is the truncation
radius of the the subhalo.
We have checked that the number of halos found in our simulations is
basically the same when using the FOF and AHF algorithms.

These two catalogues (FOF and AHF) have been used as a tool to
identify the regions of interest and to separate the contribution to
the SZ signal of galaxy clusters from that of the diffuse gas, as we
explain in \S\ref{section3}.

\section{Methodology}
\label{section3}

To characterize the statistical properties of the SZ signals associated
  with superclusters of galaxies similar to the CrB, we proceed the following
  way. 
We first identify specific regions (superclusters) within the whole MNU
simulation, which are selected to have similar characteristics (in terms of
total mass, overdensity and number density of cluster objects) to those of
CrB. All our analyses are focused on these particular regions.
Once these ``mock CrB regions'' are identified, we produce the
associated SZ maps (both thermal and kinetic components), provided
that these regions are placed at the distance of CrB, using an average
redshift for the supercluster of $z=0.07$ \citep{Ricardo2005}, and the
cosmological parameters of the simulation.

In order to isolate the contribution of the different physical
components to these maps, we produce five sets of maps, considering:
(a) all gas particles; (b) all gas particles except those which are
associated with clusters in the region; (c) only WHIM particles, defined
as those gas particles having temperatures within the range $10^5$ to
$10^7$~K; (d) only gas particles that belong to the clusters; and (e)
only gas particles that belong to groups of galaxies. The last three
sets of maps complement each other in the sense that their sum is equal to the 
map for case (a).
Finally, and in order to explore the dependence of the amplitude of
the detected signals with the relative orientation of the
supercluster, we have also produced sets of SZ maps using random
orientations of the simulation box with respect to the observer for
all the sets of maps described above.

All these steps are described in the following subsections.

\subsection{Identification of superclusters similar to CrB}

The Corona Borealis supercluster (CrB-SC) has been morphologically
described by \citet{Small97} as a \emph{flattened pancake} of $100
\runit$ a side with an estimated depth of $40\runit$.  According to
the catalogue by \citet{Einasto01}, it includes eight clusters with
total masses (including gas and dark matter) ranging from
$1.5\times10^{14}\rm M_{\odot}$ to $8.9\times10^{14}\rm M_{\odot}$,
yielding a total mass for the supercluster of $(3-8)\times10^{16}\rm
M_{\odot}$ \citep{Ricardo2005}. However, the core of the supercluster,
where the Abell clusters lie,  extends over an area of only 
$20\runit$ a side. Hence, we decided to study cubic subvolumes of
$50\runit$ a side, which were identified as similar to CrB-SC by using
the two independent criteria described below:
\begin{itemize}
\item \emph{Criterion 1: Over-densities}. Assuming that the CrB
  supercluster is spherically symmetrical and all its mass is concentrated
  in the core, its average density would be of the order of 
  $\pot{1.67}{12}\munit (\runit)^{-3}$. The average density of the 
  simulation is of the order of
  $\pot{8.33}{10}\munit (\runit)^{-3}$, calculated by dividing the
  total mass (DM + gas) between the volume of the simulation. Thus, it
  is around 20 times smaller than the CrB average density. Given this
  fact, an overdensity criterion for the gas particles in the
  simulation can be used in order to locate \emph{CrB-like} regions.

We considered as candidate regions the subvolumes of $50 \runit$ a
side that contained the largest number of overdense gas particles, i.e. 
gas particles which show a density larger than or equal to $20$ times 
the average density of the simulation. These regions are guaranteed 
to have a similar overdensity to that of CrB supercluster and enough 
gas particles within to allow further filtering when building the 
different sets of maps.
\item \emph{Criterion 2: Resemblance to CrB}. The number of clusters
  and the total mass within a given subvolume were the two
  characteristics considered in order to apply this criterion. A
  complete catalogue of the objects present in the simulation and
  their masses was required.
 
A subvolume of $50 \runit$ a side was considered as a candidate for
the study when it satisfied \textbf{1)} baryonic mass of the order of
the mass of the CrB supercluster, i.e. $M_{\rm
  gas}\geq10^{15}$~$\munit$; and \textbf{2)} the number of clusters
within is greater or equal to 6, defining a cluster as an object with
$M_{\rm Total}=M_{\rm gas}+M_{\rm
  DM}\geq5\times10^{13}$~$\munit$. Therefore, this criterion
established minimum thresholds for the total mass of the candidate
regions, the number of clusters within and the mass of these
clusters. Note that, in order to be conservative, these three
thresholds lie below those of the CrB supercluster.
\end{itemize}

\begin{table}
\centering
\caption{Sub-volumes considered in this study. Each corresponds to a
cubic box of 50~$\runit$ a side. Columns 2 and 3 indicate the gas and total
(dark+baryonic) mass within each subvolume. The last column indicates the number of
clusters identified in the box using the AHF algorithm (see text for details).}
\scriptsize{
\begin{tabular}{c c c c}    
\noalign{\smallskip}
\hline\hline
\noalign{\smallskip}
Subvol.&$M_{\rm gas}$&$M_{\rm Total}$ & No. clusters\\
 &[$\munit$]&[$\munit$ ]&$M\geq\pot{5}{13}\munit$\\
\noalign{\smallskip}
\hline
\noalign{\smallskip}
001&$2.29\times10^{15}$&$\pot{1.53}{16}$&14\\
002&$2.30\times10^{15}$&$\pot{1.54}{16}$&13\\
003&$2.01\times10^{15}$&$\pot{1.34}{16}$&11\\
004&$2.94\times10^{15}$&$\pot{1.96}{16}$&27\\
005&$1.90\times10^{15}$&$\pot{1.27}{16}$&12\\
006&$2.45\times10^{15}$&$\pot{1.63}{16}$&14\\
007&$2.85\times10^{15}$&$\pot{1.91}{16}$&19\\
008&$2.74\times10^{15}$&$\pot{1.83}{16}$&23\\
009&$2.85\times10^{15}$&$\pot{1.90}{16}$&21\\
\noalign{\smallskip}
\hline
\hline
\end{tabular}
}
\normalsize
\rm
\label{table:subvols}
\end{table}

Nine sub-volumes were finally selected and are listed in
Table~\ref{table:subvols}. All regions satisfy the two aforementioned
criteria, and are centred around the largest concentration of
over-dense particles within each of them. None of the nine sub-volumes
 overlaps another.
The last column in the table shows the total number of galaxy clusters
(defined as halos with $M \ge \pot{5}{13}$~$\munit$), which were
taken from the AHF catalogue.

\subsection{Building SZ maps}

For each gas particle $i$, the simulation provides the position
($\mathbf{r}_{\rm i}$), velocity ($\mathbf{v}_{\rm i}$), temperature ($T_{\rm
i}$), density ($\rho_{\rm i}$) and SPH smoothing length ($h_{\rm i}$).
Using the physical parameters for all the particles, we can build the SZ
maps following the procedure described in \citet{DaSilvaetal00}.
Each gas particle has an associated mass profile given by $m_{\rm gas}\
W(\mathbf{r}-\mathbf{r_{i}},h_i)$, where $\mathbf{r}$ is the coordinate vector
of the position in which we are calculating the mass profile, $\mathbf{r_{i}}$
is the coordinate vector of the position of the i-th gas particle, $h_i$ is the
smoothing length for that particle and $W$ is the normalized spherically
symmetric kernel adopted in the simulations, which in our case is given by
\citep{Springeletal01}:
\begin{equation}
\label{W_SPH}
W(x,h_i)=\dfrac{8}{\pi h_i^3}\left \{  \begin{array}{ll}
1-6x^2+6x^3, & 0\leq x\leq0.5\\
2(1-x)^3, & 0.5 < x \leq 1\\
0, & x > 1
\end{array} \right.
\end{equation}
\noindent 
where $x=|\mathbf{r}-\mathbf{r_{i}}|/h_i$. Thus, each particle occupies a sphere
with a radius equal to its SPH smoothing length.

The thermal component of the SZ effect (tSZ) in a certain direction is
given by the Comptonization $y$ parameter:
\begin{equation}
\label{yc}
y=\int n_e\dfrac{k_BT_e}{m_ec^2}\sigma_T dl
\end{equation}
where the integral is taken along the line of sight (\emph{los}),
$\sigma_T$ is the Thomson cross-section, $n_e$ is the electron number
density, $T_e$ is the electronic temperature, $k_B$ is the Boltzmann
constant and $m_ec^2$ is the electron rest mass energy.
The numerical evaluation of the integral in Eq.~\ref{yc} is done by 
discretizing the \emph{los} integral as
\begin{equation}
\label{y_approx}
y\simeq \frac{k_B\sigma_T}{m_ec^2}\frac{m_{\rm gas}}{\mu_{\rm e}m_{\rm p}}\Delta x
\sum_{\alpha}\sum_{i}T_{i}W(|\mathbf{r_{\alpha}}-\mathbf{r_{i}}|,h_i)
\end{equation}
\noindent 
where $\Delta x$ is the discrete cell size along the \emph{los}. In this
equation, the $\alpha$ index runs over all the cells along the line of
sight, the $i$ sum runs over all the particles that contribute to the
column of cells projected onto a pixel, $\mu_{\rm e}$ is the electron
mean molecular weight and $m_{\rm p}$ is the proton mass.

To build the tSZ maps, we make a further approximation, which consists
in assuming a plane--parallel projection of the simulation boxes. At
the distance of CrB, the error of using this approximation instead of
doing the exact \emph{los} integration is very small, as discussed below. 
Thus, for building the maps we create a 3D cubic grid with a cell size
of $\Delta x=80h^{-1}kpc$ a side and we use equation~\ref{y_approx} to
obtain the Comptonization parameter for each cell.

A similar method is applied for the kinetic component of the SZ
effect, given by the $b$-parameter
\begin{equation}
\label{b}
b = -\int n_e\frac{ \vpec }{c}\sigma_T dl
\end{equation}
\noindent 
where $\vpec = \vec{v} \cdot \hat{n}$ is the peculiar velocity of the cluster
projected along the line of sight ($\hat{n}$), and $c$ is the speed of
light. This equation is accordingly rewritten as
\begin{equation}
\label{b_approx}
b\simeq \frac{\sigma_T}{c}\frac{m_{\rm gas}}{\mu_{\rm e}m_{\rm p}}\Delta x
\sum_{\alpha}\sum_{i}v_{\rm p,\it i} W(|\mathbf{r_{\alpha}}-\mathbf{r_{i}}|,h_i)
\end{equation}
This process was carried out for each subvolume, yielding $y$ and $b$ maps with a pixel
size of $\Delta x=80h^{-1}kpc$. This physical length is equivalent to an angular
resolution of $1.4$~arcmin if we place the boxes at the CrB supercluster
distance (using the redshift $z=0.07$ and the cosmological parameters of the
simulation). These maps are then degraded in resolution, by a convolution with
the appropriate Gaussian beam, in order to make predictions at the
relevant angular resolutions. In particular, for the case of
  MITO/VSA, the maps are degraded to an angular resolution of $16'$.

\begin{figure}
\includegraphics[width=0.98\columnwidth]{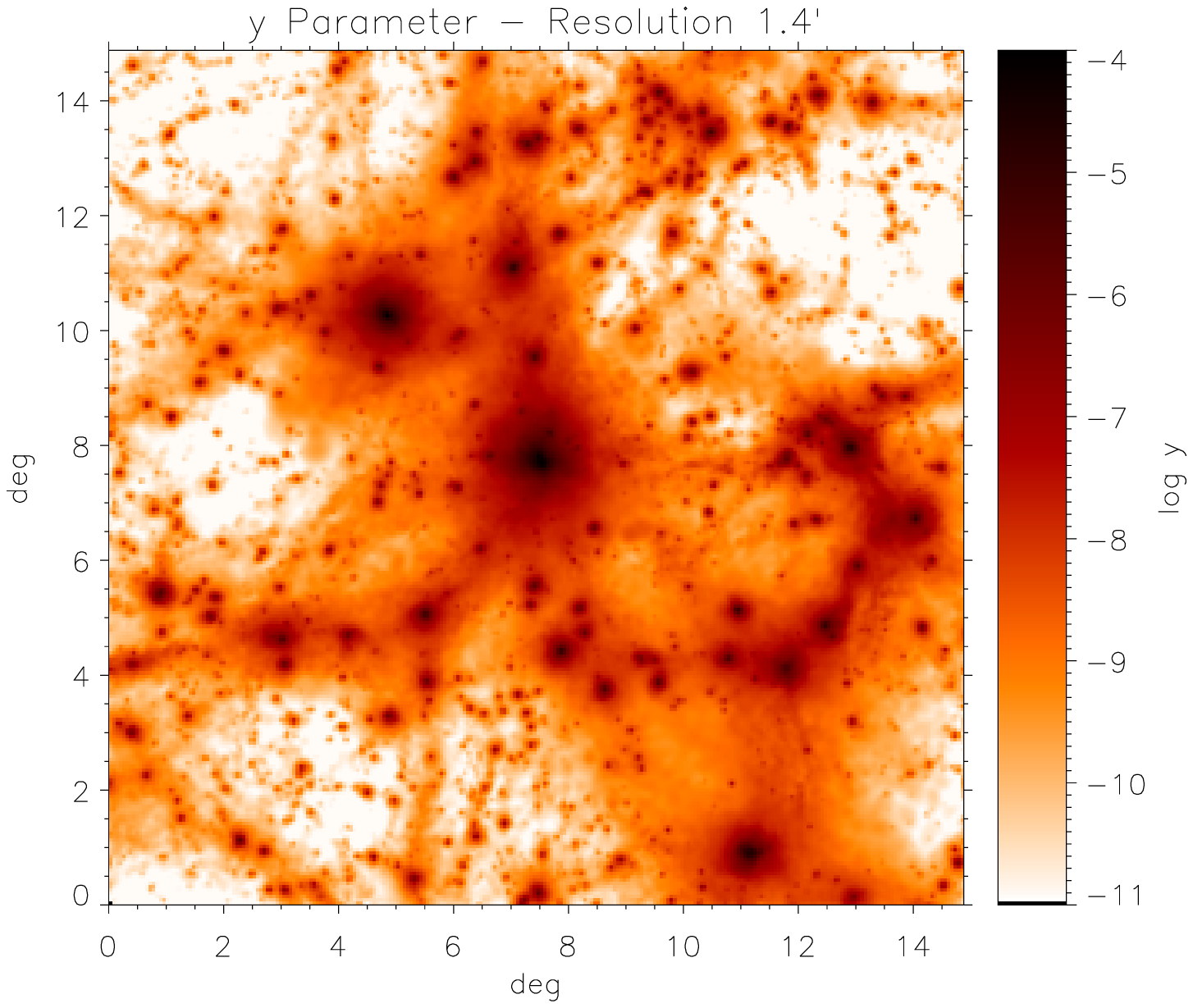}
\includegraphics[width=0.98\columnwidth]{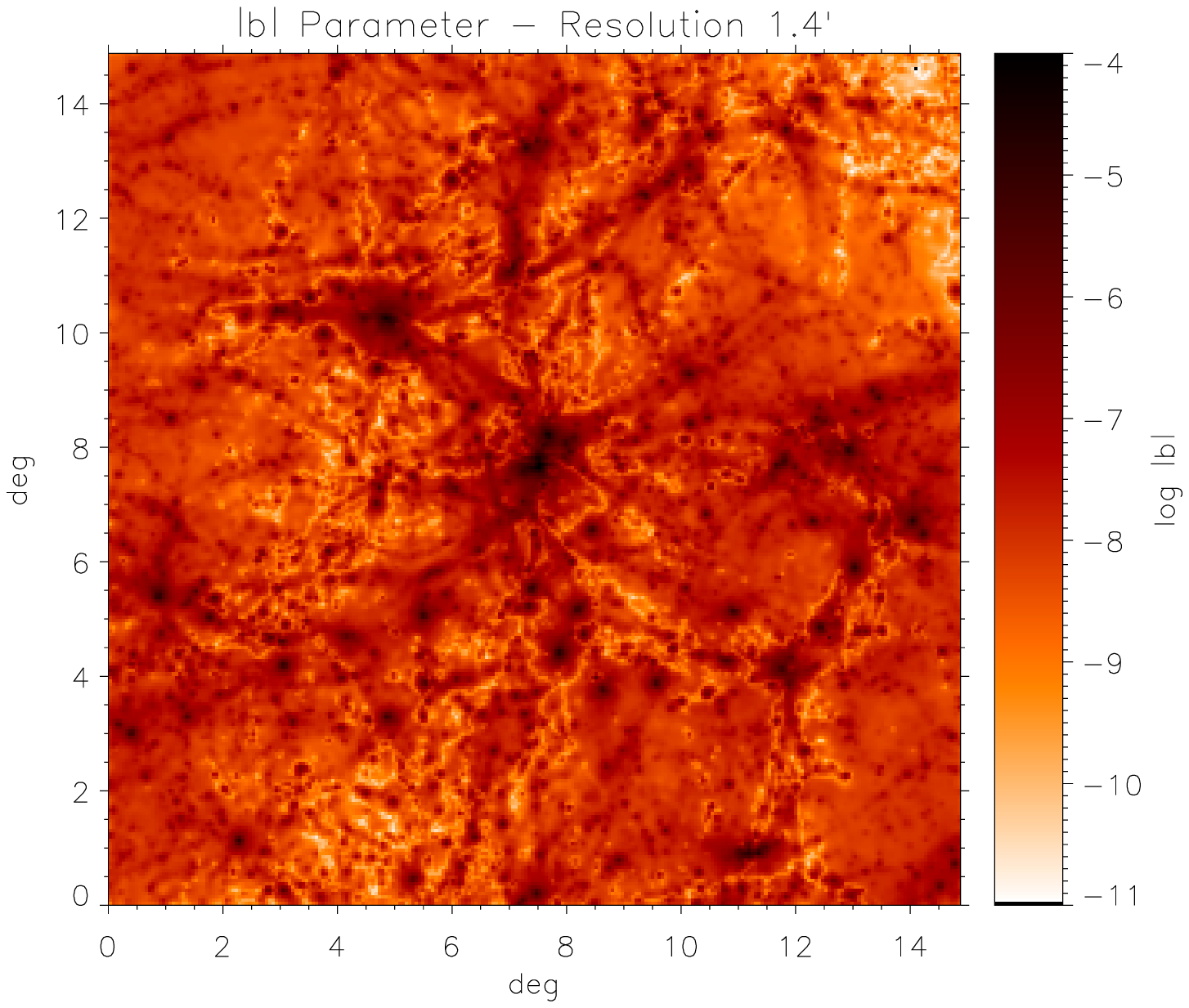}
\caption{Example of Sunyaev--Zel'dovich effect maps obtained from the
  simulations. The angular resolution of the maps is
  $1.4'$. \textbf{Top}: thermal SZ effect, given by the y-parameter map
  in logarithmic scale. \textbf{Bottom}: kinetic SZ effect, given by the
  logarithm of the absolute value of the $b$ parameter.}
\label{fig:log_both_SZ_with}
\end{figure}

\begin{figure}
\includegraphics[width=0.98\columnwidth]{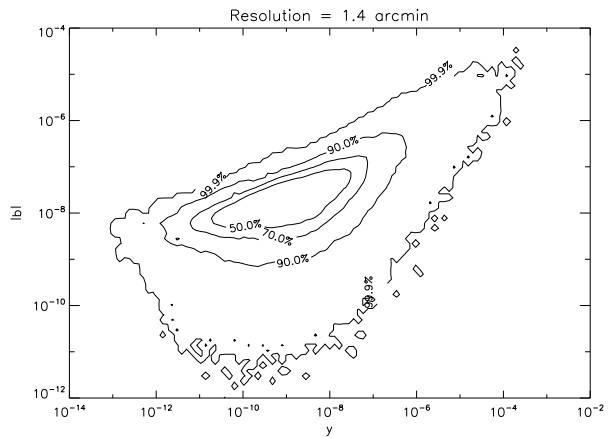}
\caption{Two-dimensional distribution of the pixel values for the set
  of nine simulated maps in supercluster regions. The two axes
  represent the $y$ Compton parameter and the (modulus of the) $b$
  parameter. Contour levels enclose the regions with 50, 70, 90 and
  99.9 per cent of the total amount of pixels, respectively. }
\label{fig:contornos_original}
\end{figure}

Examples of SZ maps for one of the subvolumes with $1.4'$ resolution are shown
in Figure~\ref{fig:log_both_SZ_with}, where it can be seen that the peak values
match those of typical clusters of galaxies, i.e. $y_{\rm max}\sim10^{-4}$ and
$|b|_{\rm max}\sim10^{-5}$. When comparing these numbers with other results from
numerical simulations \citep[e.g.][]{DaSilvaetal00,Springeletal01}, it is
important to take into account the effect of pixel scale, which in our case is
significantly larger.
Following \citet{Roncarellietal07}, Figure~\ref{fig:contornos_original} presents
the distribution of the pixel values for the whole set of nine maps, in the
plane of Doppler $|b|$ parameter versus $y$ parameter. Note that in our case, we
are integrating the SZ flux only from a $50$~$\runit$ region because we want to
isolate the effect from the superclusters, and thus the peak of this
distribution is shifted towards low values of $y$ parameter.

\begin{figure}
\centering 
\includegraphics[width=0.98\columnwidth]{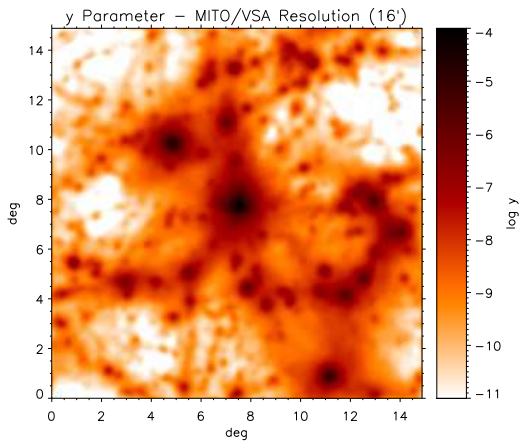}
\caption{ Comptonization parameter (on a logarithmic scale) of the same map shown
  in the top panel of figure~\ref{fig:log_both_SZ_with}, convolved with a
  Gaussian beam to match MITO/VSA resolution (16 arcmin). }
\label{fig:res_comp}
\end{figure}

Figure~\ref{fig:res_comp} shows the effect of the beam convolution on the
original map (Fig.~\ref{fig:log_both_SZ_with}), for the case of VSA/MITO
resolution ($16'$).
The average $y$ parameter in the maps with clusters and a resolution of
  $1.4'$ ranges from $\pot{0.73}{-7}$ to $\pot{1.3}{-7}$, the mean value being 
  $<y>=\pot{9.6}{-8}$.
Note that this result is roughly an order of magnitude lower than the average
value of $\left\langle y\right\rangle = \pot{2.6}{-6}$ found by
\citet{SWH2001}. The reason is again that in our case, we are only integrating a
total depth of 50~$\runit$, while in their computations they consider all the
integrated contribution up to $z=6$.

%
The approach described above makes use of two approximations: a plane--parallel
projection and approximate integral. To explore their validity, we have
analytically obtained the profile for one single particle and we compared it to
the one obtained using our map-mapping technique. This comparison yields a
point-by-point difference in the profile of less than $5\%$ when the particle is
located in the centre of the map and of less than $10\%$ when it is placed at a
distance of $\sim4\degree$ along the diagonal of the subvolume. Both the
integrated value and the maximum value show even smaller differences ($\leqslant
3\%$) when compared to the theoretical approach.  We find that, in that case, the
error in the central angular coordinate of a particle is smaller that the size
of one pixel in our integration scheme.

\subsection{Subtraction of galaxy clusters contribution to the maps}
\label{extract}

In order to characterize the SZ signal which is not associated with galaxy
clusters in the subvolumes, the contribution of those particles which belong to
clusters is eliminated from the \emph{los} integration to produce a set of SZ
maps that will contain only the contribution of WHIM and galaxy groups.
In practice, two assumptions are needed to perform these computations.  First,
we have to adopt a definition of which halos are going to be considered as
``galaxy clusters'', and thus they will be removed from the maps. In this study,
we shall define clusters as those objects more massive than $\pot{5}{13}\munit$
(in total mass). This cluster mass threshold does not significantly  alter our
results.
We note that those halos can be easily identified in any of the two catalogues
described in \S\ref{section2.1}.

Secondly, we need to specify to what extent we are removing particles from a
given halo. A reasonable assumption is to remove those particles lying within
one virial radius ($r_{\rm vir}$) from the cluster. Estimates for the virial
radius of the clusters are provided in the AHF catalogue, so we shall adopt
those values for this purpose.
Using the $r_{\rm vir}$ values adopting the mass limit of galaxies of
$M\geq5\times10^{13}\rm \munit$, clusters were extracted by using a spherical
top-hat function; i.e. neglecting the contribution of every particle that lies
within the virial radius of any given cluster (in the 3D grid).

\subsection{Rotations}

Gas in the WHIM phase is located mainly along filaments and sheet-like
structures. Because the SZ effect probes the integral contribution
of the electron pressure along the line of sight, we might expect certain
orientations in which the gas in filaments is aligned along the \emph{los} of
the observer to yield large values for the $y$ parameter, even comparable to
that of a cluster of galaxies.
In order to quantify this possibility, after subtracting the clusters, we
randomly rotated each subvolume 300~times, using rotations which were uniformly
distributed in the sphere and were defined by random triplets of the Euler
angles. Each of the three Euler angles were pseudo-randomly chosen from a
uniform distribution between $0\degree$ and $180\degree$. Using these 301~maps
(300 rotations plus the non-rotated original map) for each subvolume and each
resolution, we were able to perform a statistical study of the orientation
effects of the detected signals.

\section{Statistical Analysis of the SZ maps}
\label{section4}
Here we present an overview of the overall statistical properties of the
  simulated SZ maps, and how those properties depend on the different gas phases
  and the orientation of the supercluster.

\subsection{Analysis of the average values in the maps}

Table~\ref{table:ratios} shows the values of the ratio of total flux ($Y$) in
the tSZ maps, between the total maps and the different separated maps described
in \S\ref{section3}, for each of the nine selected subvolumes. These numbers
indicate the relative contributions of the different gas phases to the total
(integrated) y-flux in supercluster regions like CrB.
Roughly $73$~\% of the total flux in the maps comes from clusters with $M \ge
\pot{5}{13}$~$\munit$, and the other $\sim 30$~\% comes from galaxy groups
and the WHIM. Of this 30 per cent signal, roughly 60\% is coming from
groups, and 40\% is coming from the WHIM.

It is interesting to compare these values, which are associated with
superclusters, with those found in an average field. For example, in Fig.~12 in
\cite{Hallman07} it is shown that for an average field, the relative
contribution is roughly two-thirds and one-third. Thus, in the case of
superclusters, we find that the relative contribution of galaxy clusters to the
average flux is slightly larger than in an average field, as one would expect
for an overdense region.

Finally, we also show in Table~\ref{table:ratios2} the same ratios, but for
the total SZ flux, i.e. including also the kSZ component.  For this calculation,
the kSZ component is added as computed in the Rayleigh--Jeans regime
(i.e. $\Delta T_{\rm RJ} \equiv (-2y + b) T_{0}$). In this case, the total flux
is thus proportional to $\int \Delta T d\Omega$.
We note that in this case, the relative contribution of the WHIM component
increases because  at low-temperatures the kSZ contribution is
comparable in amplitude (or even dominates) over the tSZ contribution. We find
that the total SZ flux in supercluster regions is roughly explained by
two-thirds contribution of galaxy clusters, $\sim 15$\% galaxy groups and $\sim
15$\% WHIM.

\begin{table}
\caption{Ratios of total flux ($Y$) in the $y$-maps of the different
  gas phases, using MITO resolution. The cluster phase is defined by 
  objects with $M \geq\pot{5}{13}$~$\munit$. Groups are defined as 
  halos below that mass limit. The WHIM phase is defined in terms of 
  gas temperatures, as $10^5 < T < 10^7$~K. The last column shows the 
  co-added contribution of WHIM and groups of galaxies.}  
\scriptsize{
\begin{tabular}{c c c c c}    
\noalign{\smallskip}
\hline\hline
\noalign{\smallskip}
Subvol.&$\dfrac{Y_{\rm total}}{Y_{\rm clusters}}$&$\dfrac{Y_{\rm total}}{Y_{\rm groups}}$&$\dfrac{Y_{\rm total}}{Y_{\rm WHIM}}$&$\dfrac{Y_{\rm total}}{Y_{\rm non-clusters}}$\\
\noalign{\smallskip}
\hline
\noalign{\smallskip}
001 & 1.33 &  5.14 & 11.92 & 3.59 \\
002 & 1.35 &  6.14 &  9.37 & 3.71 \\
003 & 1.21 &  9.41 & 13.05 & 5.47 \\
004 & 1.41 &  6.82 &  6.68 & 3.38 \\
005 & 1.59 & 10.06 &  4.82 & 3.26 \\
006 & 1.37 &  4.56 & 15.96 & 3.55 \\
007 & 1.26 &  6.84 & 11.72 & 4.32 \\
008 & 1.43 &  4.02 & 13.00 & 3.07 \\
009 & 1.41 &  5.79 &  7.30 & 3.23 \\
\hline
Average & 1.37 & 6.53 & 10.43 & 3.73 \\
\noalign{\smallskip}
\hline
\hline
\end{tabular}
}
\normalsize
\rm
\label{table:ratios}
\end{table}

\begin{table}
\caption{Same as in Table~\ref{table:ratios}, but now for the ratio of
  total SZ flux in the maps, i.e. also including the kSZ
  component. The total SZ flux ($S$) is proportional to $\int \Delta T
  d\Omega$, where the total temperature decrement is computed as in
  the Rayleigh-Jeans regime (i.e. $\Delta T_{\rm RJ} \equiv (-2y + b)
  T_{0}$). }  
\scriptsize{
\begin{tabular}{c c c c c}    
\noalign{\smallskip}
\hline\hline
\noalign{\smallskip}
Subvol.&$\dfrac{S_{\rm total}}{S_{\rm clusters}}$&$\dfrac{S_{\rm total}}{S_{\rm groups}}$&$\dfrac{S_{\rm total}}{S_{\rm WHIM}}$&$\dfrac{S_{\rm total}}{S_{\rm non-clusters}}$\\
\noalign{\smallskip}
\hline
\noalign{\smallskip}
001 & 1.41 &  5.39 &  8.04 & 3.23 \\
002 & 1.41 &  6.65 &  6.66 & 3.33 \\
003 & 1.24 & 10.33 & 11.14 & 5.36 \\
004 & 1.52 &  8.15 &  5.11 & 3.14 \\
005 & 1.83 & 11.45 &  3.89 & 2.91 \\
006 & 1.41 &  4.77 & 10.42 & 3.27 \\
007 & 1.32 &  7.95 &  7.79 & 3.93 \\
008 & 1.58 &  4.33 &  7.34 & 2.72 \\
009 & 1.85 &  7.09 &  3.85 & 2.49 \\
\hline
Average & 1.51 &  7.35 & 7.14 & 3.38 \\
\noalign{\smallskip}
\hline
\hline
\end{tabular}
}
\normalsize
\rm
\label{table:ratios2}
\end{table}

\subsection{One-point probability distribution functions}
\label{stats}
%
A general statistical description of the maps can be given in terms of a
  fluctuation analysis \citep[see e.g.][]{RubinoSunyaev03}, which is based on
  the study of the one-point probability distribution function (1-PDF). This
  kind of analyses allows us to characterize the statistical properties of the
  sources below the confusion level in the maps. We now study separately the
  1-PDF of the thermal and kinetic SZ components.

\subsubsection{tSZ}

\begin{figure}
\centering 
\includegraphics[width=0.98\columnwidth]{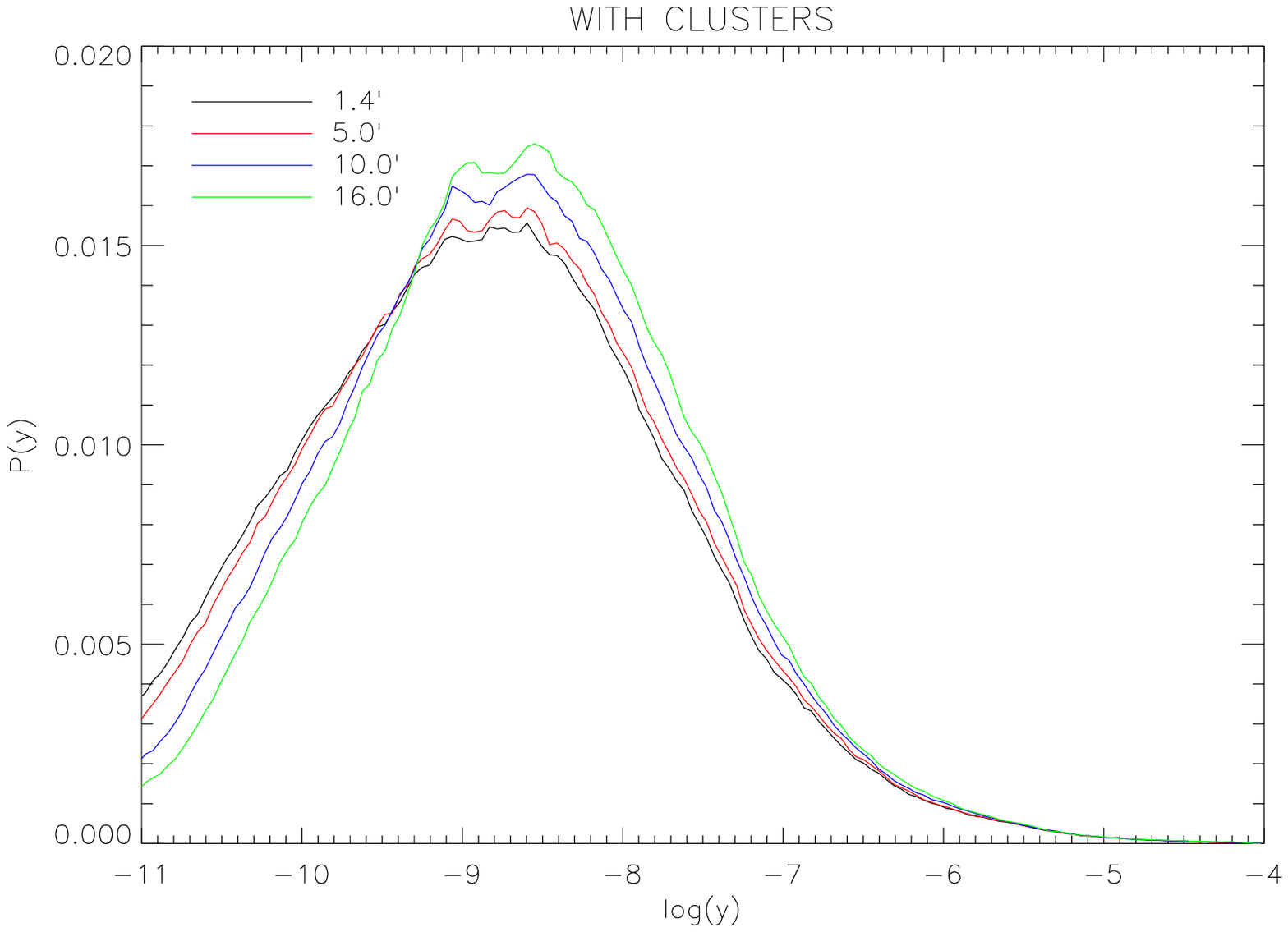}
\includegraphics[width=0.98\columnwidth]{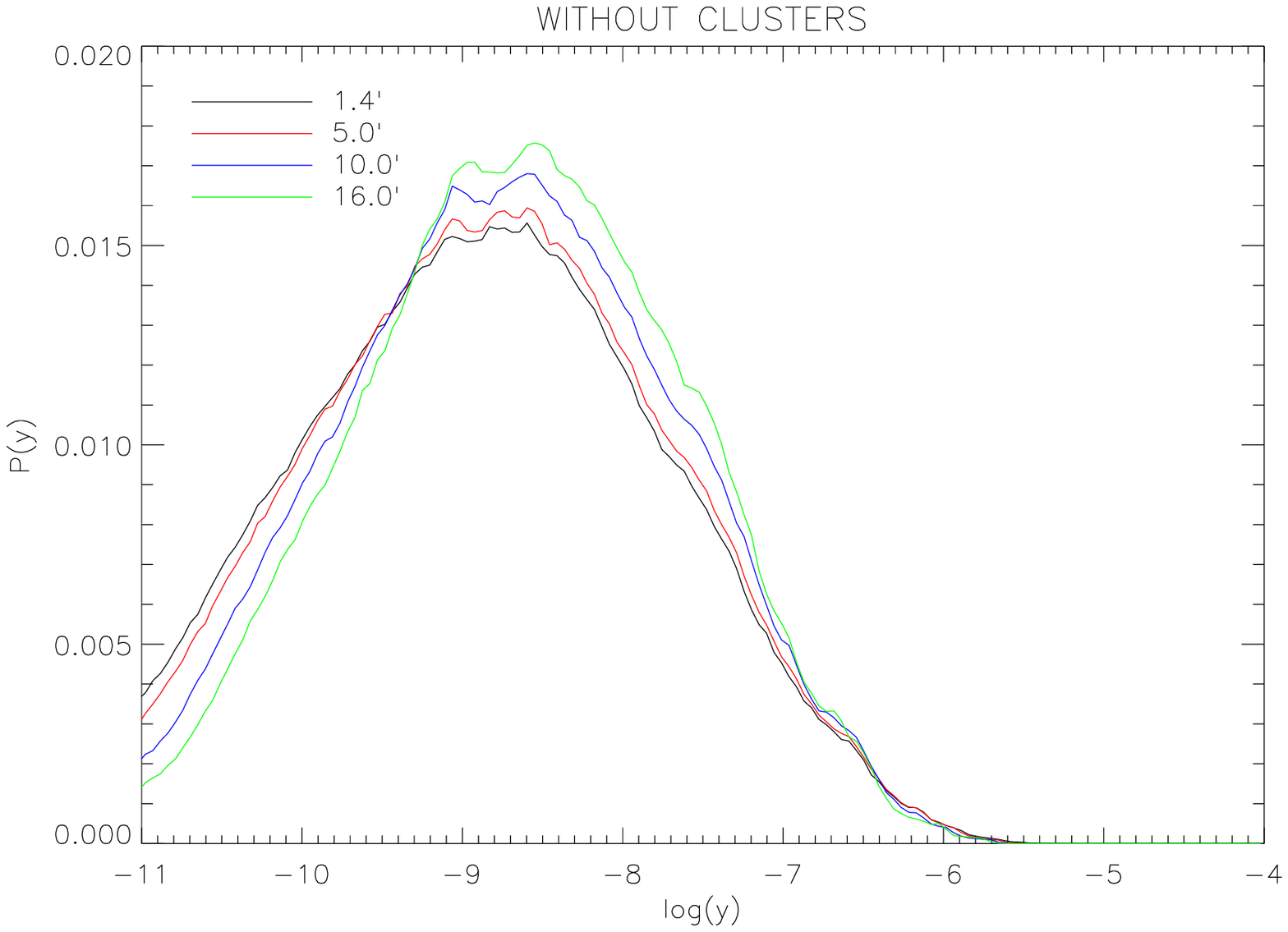}
\caption{Averaged probability distribution for tSZ ($y$ parameter). These
distributions are generated averaging over the nine subvolumes with clusters 
(top panel) and without clusters (bottom panel). The different colours correspond to 
the different resolutions used, as described in the legend. We can see that after
the cluster subtraction, the upper tail of the distribution is significantly
reduced.}
\label{fig:Probability_with_without}
\end{figure}

Fig.~\ref{fig:Probability_with_without} shows the 1-PDF for the tSZ
  component, obtained from the $y$-maps using different angular
resolutions. Comparing the two panels, it is clear that the upper tail of the
distribution is significantly reduced after cluster subtraction. However, the
tail does not disappear completely, yielding a spot with $y$ parameter values
within the $1\sigma$ region of CrB observations.

\begin{figure}
\centering 
\includegraphics[width=0.98\columnwidth]{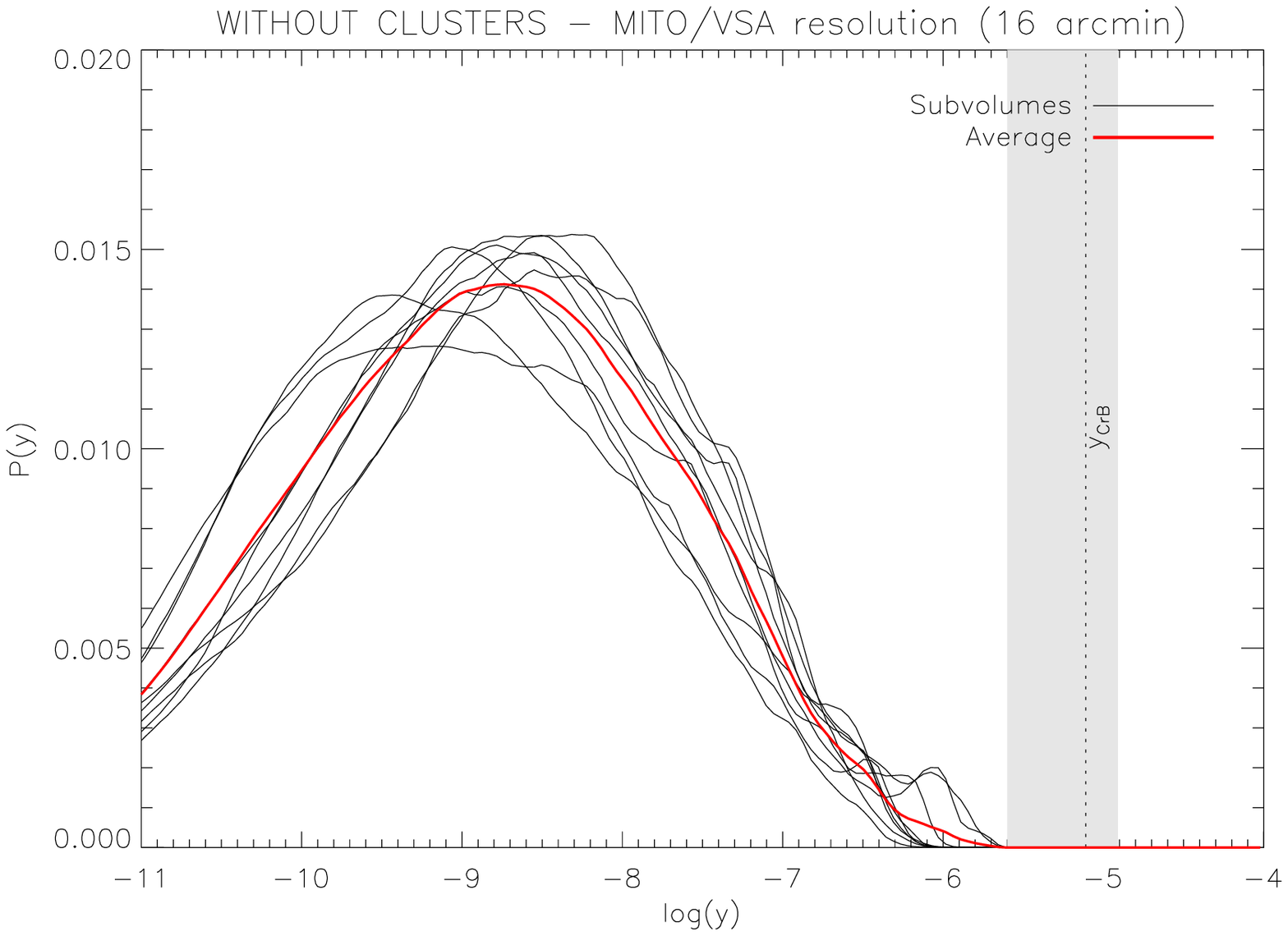}
\includegraphics[width=0.98\columnwidth]{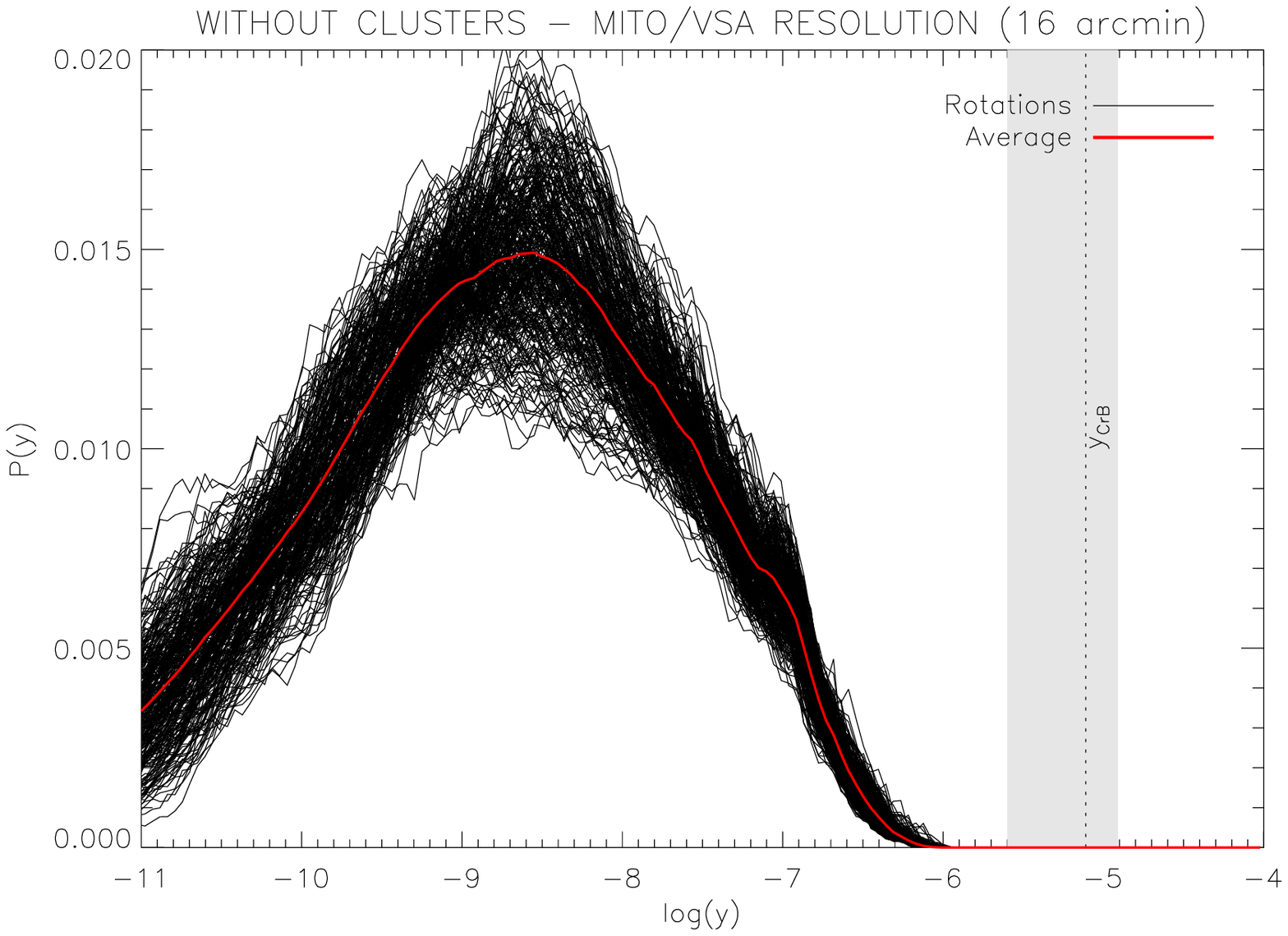}
\caption{\textbf{Top}: Averaged (over the 301 rotations) one-point probability
  distribution functions of the $y$-maps for the nine independent subvolumes
  after cluster subtraction. The red line corresponds to the average of the nine
  curves.  \textbf{Bottom}: 1-PDFs for the 301 maps obtained by rotations of a
  single subvolume. Again, the red line corresponds to the average of all black
  lines. }
\label{fig:Probability_subvols_rots}
\end{figure}

%
\paragraph{Rotations of the tSZ maps. } 
We have carried out a statistical study to characterize the dependence
  of the tSZ signal on the different orientations of the observer with respect
  to the supercluster. Our aim is to quantify how random alignments of gas
  filaments along the line of sight affect the 1-PDF $y$-distributions.
The top panel in Fig.~\ref{fig:Probability_subvols_rots} shows the nine 1-PDF
curves after averaging over the 300 rotations in each case, while the bottom
panel shows the 300 curves for a single subvolume.
Note that when studying the upper tail of the 1-PDF for a single subvolume
(bottom panel), the dispersion introduced by the rotations is very small,
suggesting that the (few) objects which are producing those signals are probably
very close to spherical. In other words, this high-$y$ tail is dominated by the
contribution of clusters below the subtraction limit, or by galaxy groups.
A very small number of possible orientations have yielded maximum values
  of the Comptonization parameter compatible at the 1~sigma level with the
  $y$-signal measured in the CrB-SC. We come back to this point in
  \S\ref{section5}.

\begin{figure}
\centering 
\includegraphics[width=0.98\columnwidth]{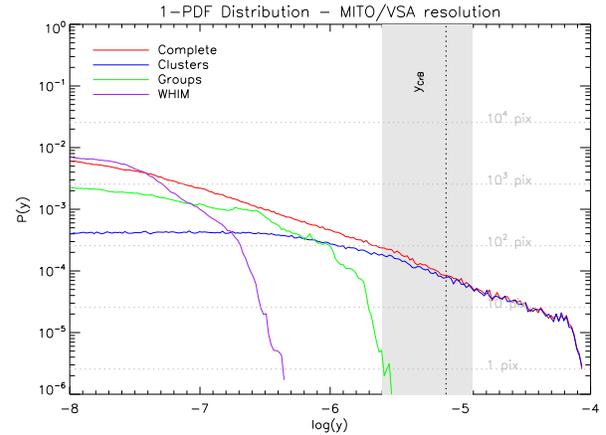}
\caption{Averaged $y$ parameter probability distribution for the nine subvolumes. 
The different colours correspond to the different components found in the maps: all 
the objects, only clusters, only groups and only WHIM. The resolution of the maps
used to compute these distributions is $16'$, the MITO/VSA resolution. The shaded 
area corresponds to the $1\sigma$ interval of the observations in the 
CrB-SC and the horizontal dashed lines show the number of pixels that matches
a given probability, showing that the WHIM fails to build up enough tSZ signal to
be responsible for the spot observed in CrB-SC.}
\label{fig:Probability_components}
\end{figure}

\paragraph{Relative contribution of the WHIM to the total tSZ signal. }
We have also performed a detailed study of the separate contributions to the
total tSZ effect from the different physical gas components: galaxy clusters,
galaxy groups and WHIM.
The different 1-PDF's associated with each case are shown in
Fig.~\ref{fig:Probability_components}, now only for the angular resolution of
MITO/VSA (16'). As expected from the densities and temperatures of the different
gas phases, galaxy clusters are responsible of the high-$y$ tail of the 1-PDF
distribution, while groups and the WHIM have relevant contributions at the lowest
$y$-values.

When comparing with the $y$-value measured in CrB-SC, these simulations show
that only compact objects (most probably galaxy clusters, although we cannot
completely discard groups of galaxies) are the only component that might build
up a sufficiently large tSZ signal.
Based on the probablility distributions obtained in our simulations,
 the  WHIM is not a likely cause of the $y$-signal detected in the CrB-SC, as its
  probability distribution reaches the zero level at $y$ values much smaller
  than those of CrB. We will investigate this issue further  in \S5.

\subsubsection{kSZ}

\begin{figure}
\centering 
\includegraphics[width=0.98\columnwidth]{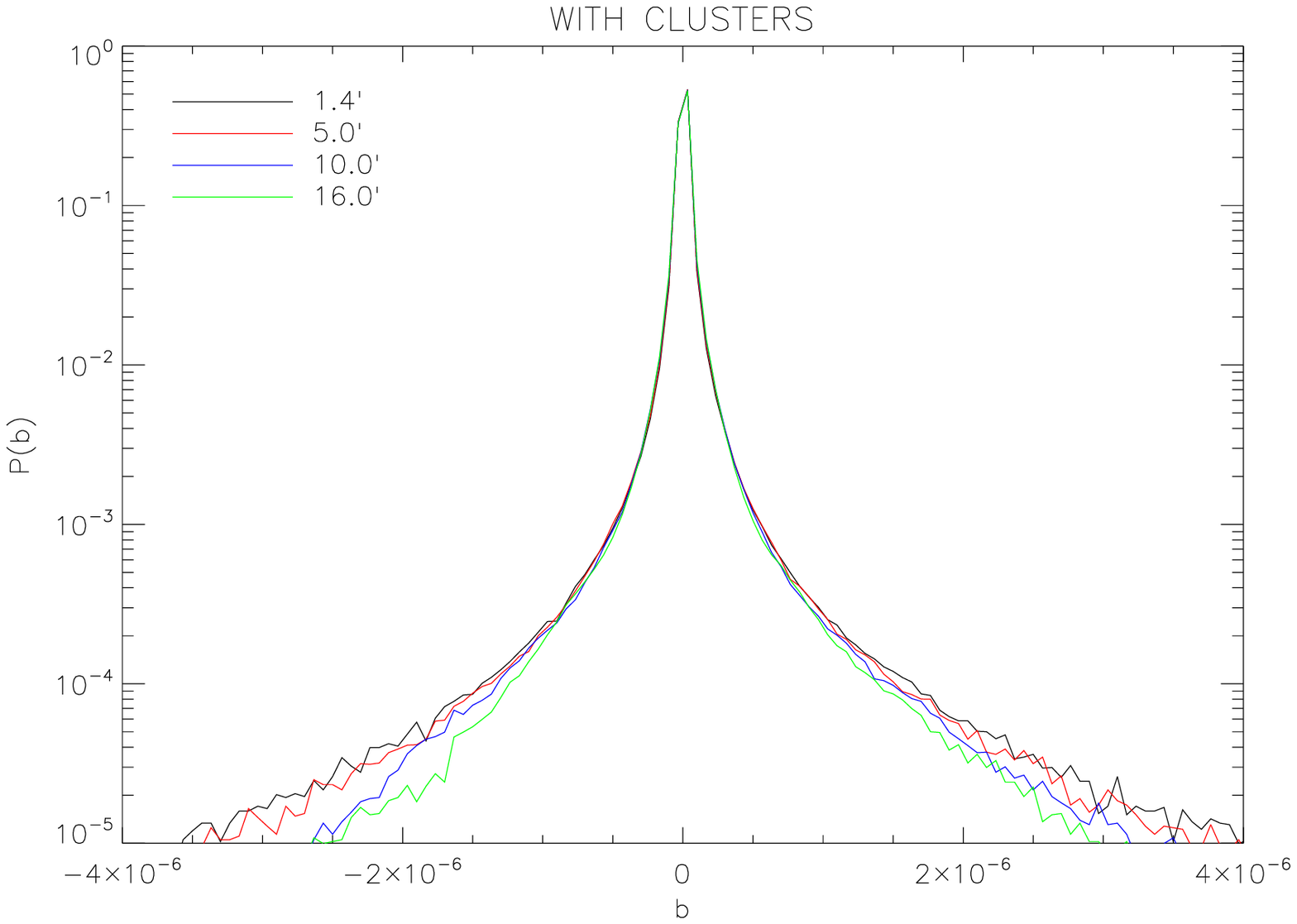}
\includegraphics[width=0.98\columnwidth]{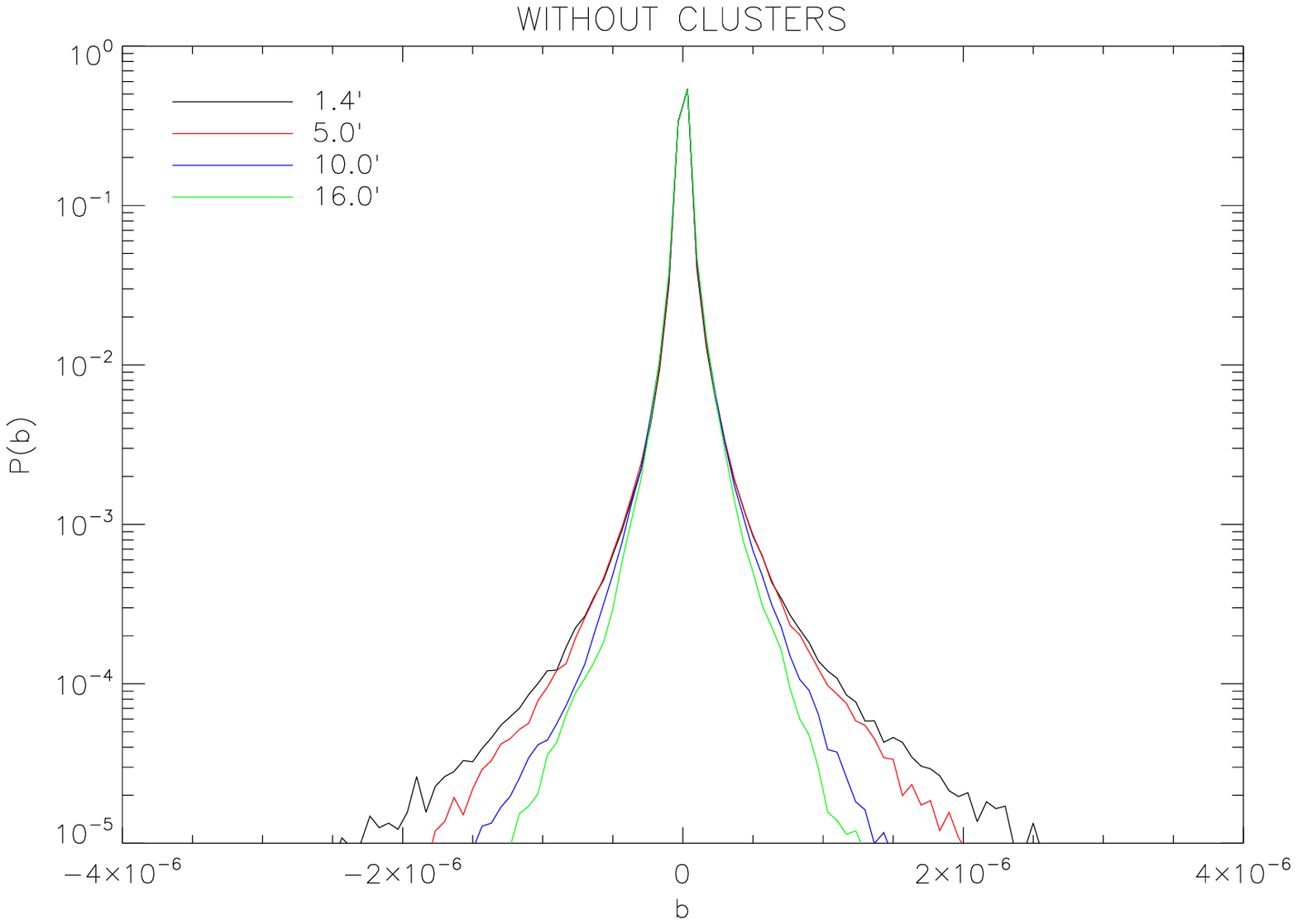}
\caption{Same as Fig.~\ref{fig:Probability_with_without}, but for the
  $b$ parameter maps (kSZ component). }
\label{fig:P_b}
\end{figure}

Figure~\ref{fig:P_b} shows the 1-PDF for the $b$ maps obtained from the nine
sub-volumes, at the same angular resolutions as in
  Figure~\ref{fig:Probability_with_without}. These distributions show that only
  clusters contribute to the high kSZ effect tails, as one would in principle
  expect given their typical bulk velocities. Considering the rotations of the
  kSZ maps does not increases the maximum kSZ signals which are detected in the
  maps.

\begin{figure}
\centering 
\includegraphics[width=0.98\columnwidth]{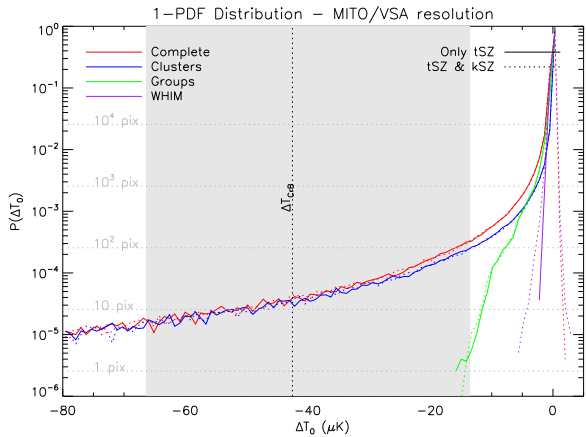}
\caption{Probability distribution of the temperature decrements
averaged over the nine different subvolumes and using MITO/VSA resolution
$16'$. Each colour shows a different component of the map, as given in the
legend; solid lines represent only the contribution to the temperature decrement
of the tSZ effect while dotted lines represent the total temperature decrement
(considering both tSZ and kSZ). The shaded area corresponds to the $1\sigma$ 
interval of the observations carried out in the CrB-SC. The horizontal dashed 
lines indicate the equivalence between 1-PDF probability and the number of pixels
in the maps.}
\label{fig:Probability_Temperature_Decrement}
\end{figure}

Finally, we have also studied the relative contribution of the kSZ component to
the total SZ signal. As we can see in
Fig.~\ref{fig:Probability_Temperature_Decrement}, the probability distribution
of the temperature decrement seems almost unaffected when considering both tSZ
and kSZ for all components except for the WHIM, in which case the probability
distribution is slightly broadened with respect to the distribution considering
tSZ only. This matches the expected result, as kSZ is only relevant to the total
temperature decrement in regions with low temperatures, so neither clusters nor
groups were expected to have a significant kSZ effect.

In any case, when considering the overall contribution of the different
  phases, it seems that the inclusion of a kSZ component cannot help in
  increasing the total SZ signal to the $-230$~$\mu$K observed with the VSA
  in the CrB-H spot. As a reference,
  Fig.~\ref{fig:Probability_Temperature_Decrement} shows as a shaded region the
  one sigma confidence interval measured by MITO/VSA ($\Delta T_{\rm CrB}^{tSZ}
  \equiv -2 y_{\rm CrB}$), and which accounts for 25\% of the total observed
  decrement.

\section{The case of Corona Borealis}
\label{section5}
In the previous section, all the statistical analyses were focused on
  the description of global properties of the SZ signals associated with
  supercluster regions. We now restrict ourselves to the particular case of the
  detection of a significant negative feature in the Corona Borealis
  supercluster (CrB-SC) with the combined MITO/VSA experiments. The goal is 
twofold: on the one hand, we want to probe if the MNU simulations are able to
  explain the detected tSZ signal by MITO/VSA in terms of gas associated with the
  supercluster. On the other hand, we want to explore if the kSZ signal
  associated with the supercluster can be a plausible explanation for the rest of
  the observed decrement in the VSA map (a signal of the order of
  $-230$~$\mu$K). As pointed out above, even when we remove the observed tSZ
  component from the total VSA decrement, the remaining signal (roughly 75\% of
  the total) is still difficult to reconcile with a primordial Gaussian CMB
  spot.

Given that the details of the sky coverage may introduce biases in the
  results, throughout this section we restrict our analyses to a sky coverage
  equivalent to the original VSA survey in which the cold spot was discovered,
  i.e. $\sim 24$~deg$^2$ \citep[see][for details]{Ricardo2005}.  In practice,
  this means that all our analyses in each of the synthetic maps are restricted
  to a circle with an equivalent area of $24$~deg$^2$, centred on the central
  pixel of each map.

As shown in the previous section, the overall analysis in all the maps
  showed that the WHIM cannot provide a high enough tSZ signal to be the sole cause
  of the $y$-signal observed in the CrB-SC.
When we include the contribution of groups of galaxies and small galaxy clusters
(below the mass threshold for extraction), we cannot rule out completely the
possibility that we might indeed be facing the cause of the observed
$y$-signal. On the other hand, when considering the kSZ contribution,
  the total decrement that can be produced in this case is still far from the
  observed cold spot in the VSA map ($-230$~$\mu$K).

In order to quantify how likely groups and small clusters could yield
  tSZ signals in the confidence interval of the observations, and how large 
  the contribution of the kSZ component to the total signal is, we have analysed
  the maximum amplitudes obtained in the simulated maps.
The top panel of Table~\ref{table:max_maps} shows these results for the case of the full
  maps (no cluster subtraction applied), while the bottom panel of the same table
  presents the case of the maps after the cluster subtraction (hence including
  groups and small clusters of galaxies). The three columns show the maximum
  values for the  $y$ and $|b|$ parameters, and the largest temperature decrement in
  the Rayleigh--Jeans regime, computed as $\Delta T_{\rm RJ} \equiv (-2y + b)
  T_{0}$.
The values in Table~\ref{table:max_maps} are in good agreement with the typical
peak amplitudes expected for the galaxy clusters in CrB, based on their X-ray
luminosity (see the estimated values in \cite{Ricardo2005}, based on the scaling
relation between X-ray luminosity and tSZ peak temperature by
\cite{2004MNRAS.347..403H}).

The values from the bottom panel of Table~\ref{table:max_maps} are significantly 
  smaller than those from the original maps, and we find that none of the subvolumes has
  maximum values within the $\pm1\sigma$ range of the $y$-value observed in the
  CrB supercluster.
If we consider a wider range, we find four subvolumes within the $\pm1.3\sigma$
region of the $y$-signal, and all of them lie within the $\pm1.36\sigma$
interval around the Comptonization parameter found in the CrB-SC observations.

That table also incorporates the maximum amplitudes (in absolute
  values) which are detected in the kSZ maps. While in the maps with clusters $y$
  is about one order of magnitude larger than $b$, in the maps without clusters,
  both components of the SZE are comparable in amplitude, thus producing a
  larger temperature decrement in those cases in which the kSZ component
  contributes with a negative sign (i.e. a clump moving away from the
  observer). In any case, the addition of the kSZ cannot provide a significant
  contribution to the total temperature decrement observed by the VSA
  ($-230$~$\mu$K).

\begin{table}
\begin{center}
\caption{Extreme values of SZ effect in the maps using MITO/VSA resolution. The top panel 
shows the results for the maps with clusters while the bottom panel corresponds to the 
cluster subtracted maps. Temperature values in the last column are obtained for the 
Rayleigh--Jeans region, in which we have $\Delta T_{\rm RJ} \equiv (-2y + b) T_{0}$.}
\scriptsize{
\begin{tabular}{c c c c}    
\noalign{\smallskip}
\hline\hline
\noalign{\smallskip}
 &\multicolumn{3}{c}{Maps with clusters}\\
\noalign{\smallskip}
Subvol. \# &$\max(y)$&$\max(|b|)$&$\Delta T_{\rm RJ}$\\
\noalign{\smallskip}
 &$[10^{-6}]$&$[10^{-6}]$&$[\mu K]$\\
\hline
\noalign{\smallskip}
001&$67.4$&$2.90$&$-372.60$\\
002&$44.8$&$2.40$&$-244.91$\\
003&$82.0$&$7.64$&$-426.56$\\
004&$40.7$&$5.77$&$-207.81$\\
005&$16.4$&$2.09$&$-86.86 $\\
006&$77.2$&$3.13$&$-418.34$\\
007&$87.7$&$6.48$&$-464.30$\\
008&$71.3$&$2.94$&$-394.63$\\
009&$30.9$&$6.62$&$-182.62$\\
\noalign{\smallskip}
\hline\hline
\noalign{\smallskip}
 &\multicolumn{3}{c}{Maps with clusters subtracted}\\
\noalign{\smallskip}
Subvol. \# &$\max(y)$&$\max(|b|)$&$\Delta T_{\rm RJ}$\\
\noalign{\smallskip}
 &$[10^{-6}]$&$[10^{-6}]$&$[\mu K]$\\
\hline
\noalign{\smallskip}
001&$1.05$&$1.31$&$-7.03 $\\
002&$0.631$&$1.34$&$-5.64 $\\
003&$0.613$&$0.995$&$-4.18 $\\
004&$0.682$&$2.24$&$-2.30 $\\
005&$0.886$&$1.21$&$-8.11 $\\
006&$2.17$&$1.35$&$-11.60$\\
007&$0.618$&$0.497$&$-2.50 $\\
008&$2.01$&$0.489$&$-11.44$\\
009&$0.691$&$0.763$&$-4.37 $\\
\noalign{\smallskip}
\hline\hline
\label{table:max_maps}
\end{tabular}
}
\normalsize
\rm
\end{center}
\end{table}

\subsection{Rotations} 
Here we also investigate how random alignments of matter along the line
  of sight might affect the maximum detectable SZ signal.
The maximum values of tSZ, kSZ and total temperature decrement in the
  Rayleigh--Jeans regime, considering all 301 maps for each subvolume are
  summarized in the top section of Table~\ref{table:summary_rots}. We also
  include the ranges of variation of these quantities (i.e. the  difference between
  the highest and the lowest maximum signal found in any of the rotations), and
  the number of rotations in which the maps show maximum values of the
  Comptonization parameter compatible at the 1~sigma and 1.3~sigma levels with the
  observed $y$-signal in the CrB-SC. Note that when looking at the total
  temperature decrement we never obtain a value as high as the $-230$~$\mu$K
  observed for the full decrement by the VSA. Therefore, for the rest of this
  subsection we focus our attention on the tSZ component.

\begin{figure}
\centering 
\includegraphics[width=0.98\columnwidth]{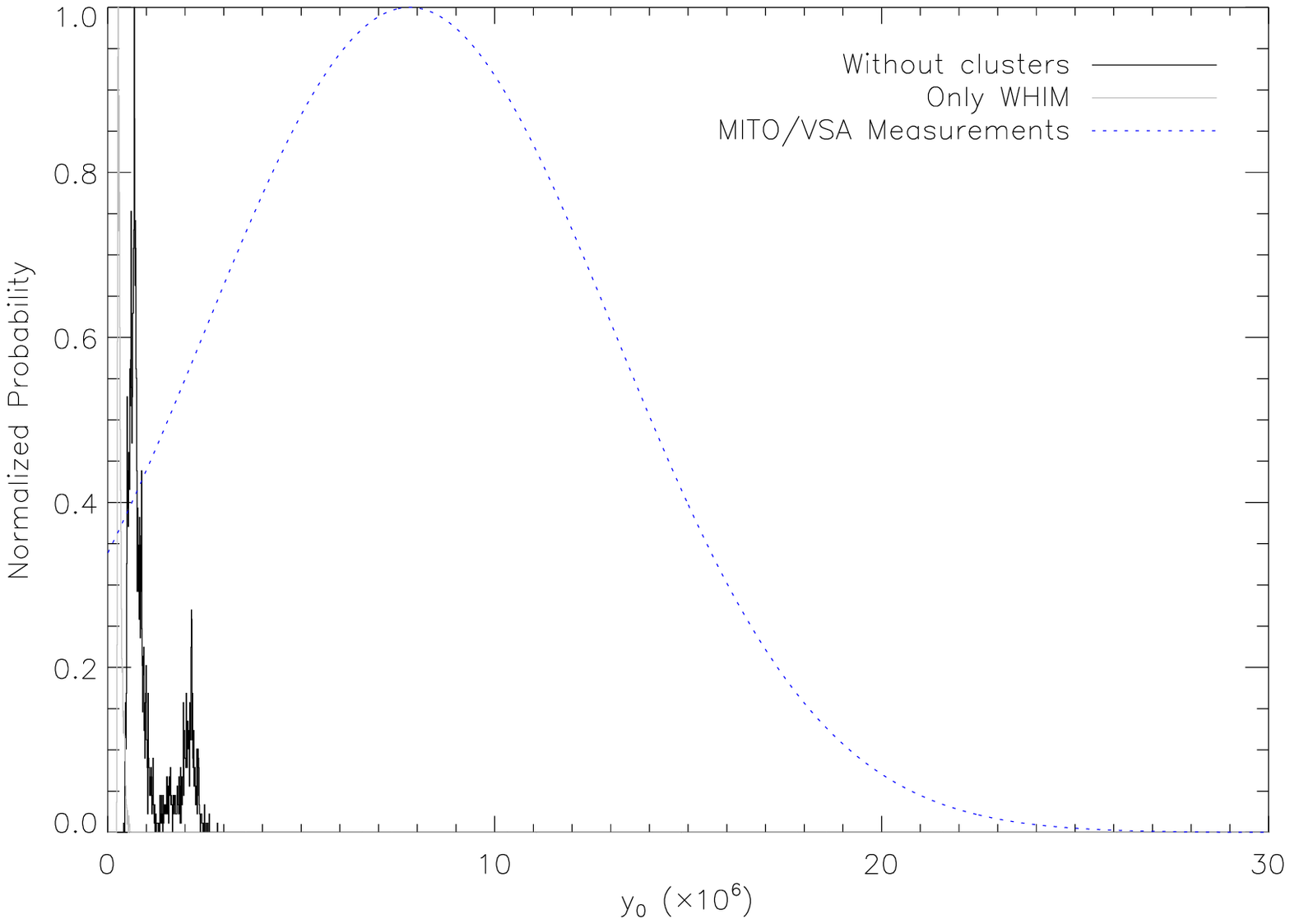}
\includegraphics[width=0.98\columnwidth]{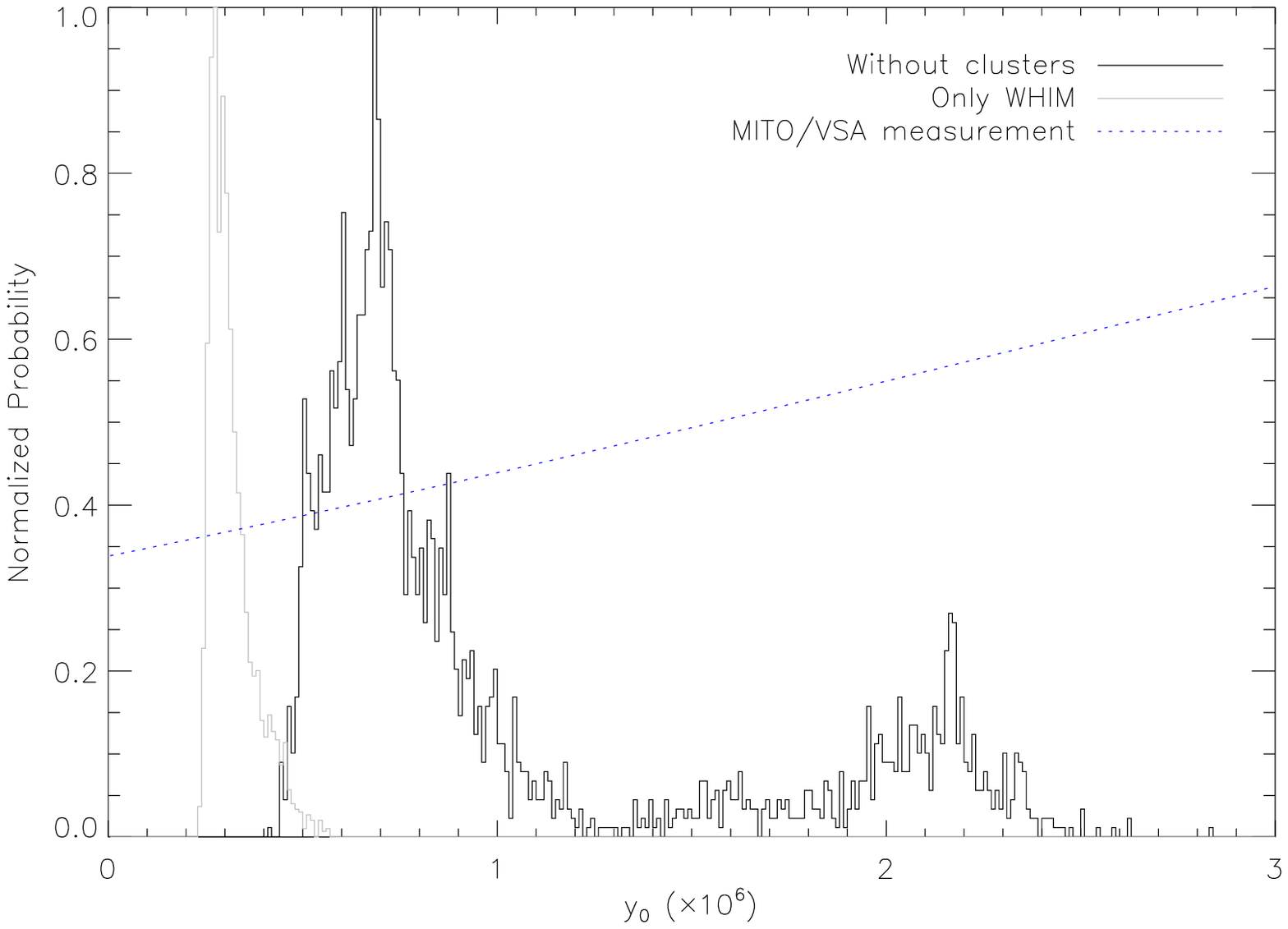}
\caption{Top: Normalized probability distributions of the
    maximum values of the $y$ parameter. Black solid line shows this
    distribution for the maps without clusters (considering all 2709
    maps; 301 maps per each subvolume), gray solid line shows the
    distribution for the maps that only include WHIM. Finally, the
    dotted line shows the posterior distribution of the SZ
    observations of the spot obtained in CrB with MITO/VSA. All
    distributions have been normalized to unity at their peak. Bottom:
    A zoom of the above plot around the region $[0,\pot{3}{-6}]$ in
    comptonization parameter.}
\label{fig:P_y_max}
\end{figure}

The full set of rotated $y$-maps also allows us to compute a crude estimate
  of the probability distribution function for the expected maximum values of
  the $y$ parameter at this resolution (MITO/VSA) and sky coverage
  ($24$~deg$^2$), which is shown in Figure~\ref{fig:P_y_max}. Although the
  features in this curve are dominated by the fact that there are only 9
  fully independent sub-volumes in the computation\footnote{The peak centered around
$y\simeq\pot{2.1}{-6}$ is caused by large groups of galaxies (with
masses just below the mass threshold used to define clusters) in
subvolumes 6 and 8.}, this figure summarizes in a
  single plot the full range of $y$-values presented in
  Table~\ref{table:summary_rots}.

The first conclusion is that even when selecting a
  ``privileged'' orientation of the supercluster with respect to the observer,
  the expected tSZ signals are still far from the MITO/VSA measurement obtained
  in CrB. Less than 0.3\% of the rotations show values compatible at 1~sigma
  with the observations.
Moreover, the range of variation of the maximum signals is very small (of the
order of $50\%$ of the maximum signal), suggesting that the source of the high
tSZ signals are compact and nearly spherical objects (most likely groups or
small clusters of galaxies).

We again emphasize that these rotations only show the effect of the
  alignment of matter that belongs to the supercluster. Hence, the ranges of
  variation shown in table~\ref{table:summary_rots} for each subvolume are only
  due to gas within $50\runit$, either in groups of galaxies or in a diffuse
  phase.

Although the previous results firmly suggest that a $y$-signal like the
  one observed is unlikely caused exclusively by WHIM in the supercluster, we
  have again subjected the WHIM-only subvolumes to random rotations in order to
  verify this statement. A summary of the results is shown in the bottom part of
  Table~\ref{table:summary_rots}, where we can see that the maximum tSZ effect
  caused by WHIM is $y \le \pot{5.6}{-7}$, hence, one order of magnitude smaller
  than the observations in the CrB-SC. The total temperature decrement also
  falls far from the observed value, always $\Delta T>-7.3$~$\mu$K, which is
  around 14 times smaller than $y_{\rm CrB}$. None of the randomly rotated WHIM
  maps yield SZ signals within the $1\sigma$ interval of the observations.
Moreover, this contribution to the total tSZ signal would be masked by the
contribution to the tSZ line of sight integral by the background of the region,
which would be of the order of $<y> \approx \pot{2}{-6}$ \citep[see
  e.g.][]{Springeletal01, Roncarellietal07}.

Hence, the second conclusion is that according to these simulations, the
  WHIM is not a likely cause of the spot, even at ``privileged 
  orientations``, where matter is aligned with the observer.

However, these conclusions do not take into account that the
observations yielded a very broad posterior distribution, as seen in
Fig.~\ref{fig:P_y_max}. Hence, we need to use a probability estimator
that weights the probability distribution for the maximum values
($y_{\max}$) that we obtain in the simulation with this broad
distribution. This probability is defined as follows:
\begin{equation}
\label{prob}
P = \int_{0}^{\infty} \Big( \int_{y}^{\infty} P_{\rm s}(y_{\max})dy_{\max} \Big) P_{\rm obs}(y) dy
\end{equation}
\noindent where $P_{\rm s}$ stands for the probability
distribution for the $y_{\max}$ values in superclusters in the
simulations and $P_{\rm obs}$ is the observed probability distribution
for the spot. For this computation, we use the curves plotted in
Fig.~\ref{fig:P_y_max} (properly normalized to get a unity integral)
as an estimate for $P_{\rm s}$, while $P_{\rm obs}$ is assumed to have
a Gaussian shape with a sigma of $\pot{5.3}{-6}$, taken from the
MITO/VSA observations. The integration of Eq.~\ref{prob} gives a
probability of $P=3.2\%$ for the maps without clusters (which still
include groups) and a probability of $P=0.9\%$ for the maps that only
include WHIM. Hence, supporting our conclusion that neither groups of
galaxies, nor WHIM are likely explanations for the tSZ component of
the spot.

\begin{table*}
\begin{center}
\caption{Orientation effects on the extreme values of the SZ effect, using the resolution
of MITO/VSA in the maps. The top panel shows the results for cluster-subtracted maps (and hence
includes the contribution of both the WHIM and small groups of galaxies). The bottom panel shows
the results for the maps only considering the WHIM contribution. The columns show: 
(1) subvolume identification number; 
(2) maximum value of the $y$ parameter considering every map of the given subvolume, in units of $10^{-6}$; 
(3) minimum value of $y_{\rm max}$ obtained after rotations, in units of $10^{-6}$; 
(4) maximum of the absolute value of the $b$ parameter considering every map of the given subvolume, in units of $10^{-6}$; 
(5) minimum value of $|b|_{\rm max}$ obtained after rotations, in units of $10^{-6}$
(6) largest temperature decrement $\Delta T_{\rm min}$, expressed in $\mu K$ in the Rayleigh-Jeans regime, considering every map
 of the given subvolume. N.B. Temperature decrements are negative quantities, hence shown here are the minimum values obtained;
(7) minimum temperature decrement in the Rayleigh-Jeans regime obtained after rotations, expressed in $\mu K$;
(8) number of independent maps that yield maximum values of the $y$ parameter in the $\pm$1sigma range of CrB.
(9) number of independent maps that yield maximum values of the $y$ parameter in the $\pm$1.3sigma range of CrB.}
\scriptsize{
\begin{tabular*}{0.8\textwidth}{c c c c c c c c c}    
\noalign{\smallskip}
\hline\hline
\noalign{\smallskip}
\multicolumn{9}{c}{Maps with clusters subtracted (Only galaxy groups and WHIM)}\\
\noalign{\smallskip}
Subvol.&$\max(y_{\rm max})$&$\min(y_{\rm max})$&$\max(|b|_{\rm max})$&$\min(|b|_{\rm max})$&$\min(\Delta T_{\rm min})$&$\max(\Delta T_{\rm min})$&$N_{\pm 1\sigma}$&$N_{\pm 1.3\sigma}$\\
\noalign{\smallskip}
&$[10^{-6}]$&$[10^{-6}]$&$[10^{-6}]$&$[10^{-6}]$&$[\mu K]$&$[\mu K]$& & \\
\hline
\noalign{\smallskip}
001&$1.39$&$0.77$&$1.80$&$0.68$&$-11.05$&$-5.64$&$0$&$193$\\
002&$1.30$&$0.42$&$2.08$&$0.83$&$-8.83 $&$-3.62$&$0$&$19 $\\
003&$1.08$&$0.54$&$1.48$&$0.91$&$-6.23 $&$-3.22$&$0$&$7  $\\
004&$1.04$&$0.58$&$2.65$&$1.46$&$-3.80 $&$-1.56$&$0$&$19 $\\
005&$1.07$&$0.49$&$1.60$&$0.91$&$-10.22$&$-5.44$&$0$&$8  $\\
006&$2.62$&$1.25$&$1.35$&$0.42$&$-14.63$&$-6.84$&$5$&$301$\\
007&$1.01$&$0.44$&$1.89$&$0.44$&$-5.00 $&$-1.76$&$0$&$18 $\\
008&$2.84$&$1.57$&$1.03$&$0.40$&$-17.22$&$-8.94$&$3$&$301$\\
009&$1.05$&$0.60$&$2.57$&$0.62$&$-12.72$&$-4.17$&$0$&$4  $\\
\noalign{\smallskip}
\hline\hline
\noalign{\smallskip}
\multicolumn{9}{c}{Only WHIM}\\
\noalign{\smallskip}
Subvol.&$\max(y_{\rm max})$&$\min(y_{\rm max})$&$\max(|b|_{\rm max})$&$\min(|b|_{\rm max})$&$\min(\Delta T_{\rm min})$&$\max(\Delta T_{\rm min})$&$N_{\pm 1\sigma}$&$N_{\pm 1.3\sigma}$\\
\noalign{\smallskip}
&$[10^{-6}]$&$[10^{-6}]$&$[10^{-6}]$&$[10^{-6}]$&$[\mu K]$&$[\mu K]$& & \\
\hline
\noalign{\smallskip}
001&$0.51$&$0.22$&$0.91$&$0.36$&$-3.91$&$-1.51$&$0$&$0$\\
002&$0.53$&$0.19$&$1.57$&$0.75$&$-6.25$&$-3.11$&$0$&$0$\\
003&$0.50$&$0.25$&$1.49$&$0.64$&$-4.55$&$-2.01$&$0$&$0$\\
004&$0.55$&$0.21$&$2.25$&$1.07$&$-2.39$&$-0.93$&$0$&$0$\\
005&$0.49$&$0.24$&$1.10$&$0.49$&$-4.84$&$-3.01$&$0$&$0$\\
006&$0.42$&$0.15$&$1.15$&$0.46$&$-4.05$&$-0.82$&$0$&$0$\\
007&$0.36$&$0.15$&$1.19$&$0.31$&$-3.28$&$-1.01$&$0$&$0$\\
008&$0.46$&$0.21$&$0.82$&$0.29$&$-3.49$&$-1.38$&$0$&$0$\\
009&$0.48$&$0.10$&$1.73$&$0.43$&$-7.32$&$-1.61$&$0$&$0$\\
\noalign{\smallskip}
\hline\hline
\label{table:summary_rots}
\end{tabular*}
}
\normalsize
\rm
\end{center}
\end{table*}

\section{Observability of diffuse SZ signals from superclusters with Planck}
\label{section6}
Although this was not the main motivation of this paper, the same set
  of MNU simulations can be used to predict the expected level of SZ signals in
  superclusters of galaxies that could be detectable with the Planck satellite.  To
  this end, we have convolved all the maps to a common (nominal) resolution of
  $5$~arcmin and we present here a brief discussion of this topic by repeating
  the same kind of analysis that was done in previous section.

We concentrate here on ``diffuse'' SZ signals associated with
  superclusters, where for the purposes of this paper ``diffuse'' means ``all
  gas phases which contribute to the SZ signal but excluding all clusters with
  masses $M \ge \pot{5}{13}$~$\munit$'' (i.e. our maps with clusters
  subtracted). A detailed study of the contribution of unresolved galaxy
  clusters (i.e. clusters with fluxes below the detection threshold for Planck,
  but with masses greater than $\pot{5}{13}$~$\munit$) is beyond the scope of
  this paper.
   
Table~\ref{table:SZ_with_Planck} shows the expected range of variation
  for the maximum SZ signals (both tSZ and kSZ) originating in superclusters of
  galaxies without considering the contribution of the clusters themselves. The top
  panel includes the contribution of groups of galaxies and diffuse gas while the
  bottom panel shows only the WHIM contribution.
Taking into account the nominal sensitivities of the Planck satellite
  (e.g. 6~$\mu$K at 143~GHz with 7.1~arcmin beam, or 13~$\mu$K at 217~GHz with a
  5~arcmin beam), direct detections of individual features due to groups+WHIM
  signals might be feasible provided that a good correction of the rest of the
  foreground components is achieved. However, a direct detection of an
  individual feature due to WHIM alone seems more difficult, although in
  principle it would be possible at the 2 or 3 sigma level.

Finally, and for illustrative purposes only, we make here a simple
  comparison with the extrapolated CrB-SC $y$-parameter value, which was measured
  using MITO/VSA. To this end, we consider two reference values, namely the peak
  of the $y$-parameter, and the minus one-sigma value, and we re-scale those two
  numbers by assuming that scale factor is the one for a point-like object. In
  this case, we obtain $\pot{8.0}{-5}$ and $\pot{2.6}{-5}$ for the central and
  minus one-sigma values, respectively. We note that those values are an order
  of magnitude larger than the maximum tSZ signals obtained in the
  maps without clusters. Hence, if we consider the Planck resolution, the
  numerical simulations are not able to explain tSZ signals as high as the one
  measured in CrB-SC (provided that the object producing it is point-like) in
  terms of hot gas in the supercluster which is not in clusters of galaxies. A
  more detailed study would require the modelling of the spot shape, which is
  beyond the scope of this article.

\begin{table*}
\begin{center}
\caption{Forecast of the maximum expected values for the diffuse SZ signals in
  supercluster of galaxies for Planck. For these computations, we adopt an
  angular resolution of 5~arcmin. For the meaning of the different columns, see
  Table~\ref{table:summary_rots} for comparison.}  \scriptsize{
\begin{tabular*}{0.68\textwidth}{c c c c c c c}    
\noalign{\smallskip}
\hline\hline
\noalign{\smallskip}
\multicolumn{7}{c}{Maps with clusters subtracted (Only galaxy groups and WHIM)}\\
\noalign{\smallskip}
Subvol.&$\max(y_{\rm max})$&$\min(y_{\rm max})$&$\max(|b|_{\rm max})$&$\min(|b|_{\rm max})$&$\min(\Delta T_{\rm min})$&$\max(\Delta T_{\rm min})$\\
\noalign{\smallskip}
&$[10^{-6}]$&$[10^{-6}]$&$[10^{-6}]$&$[10^{-6}]$&$[\mu K]$&$[\mu K]$\\
\hline
\noalign{\smallskip}
001&$3.07$&$1.56$&$7.25$&$3.90$&$-36.50$&$-19.67$\\
002&$2.91$&$0.94$&$6.05$&$2.98$&$-26.48$&$-13.05$\\
003&$4.26$&$2.02$&$4.62$&$3.68$&$-19.52$&$-9.82 $\\
004&$3.36$&$1.42$&$8.89$&$4.69$&$-13.85$&$-4.18 $\\
005&$3.55$&$2.32$&$5.78$&$4.14$&$-34.73$&$-24.03$\\
006&$4.31$&$1.74$&$3.26$&$1.41$&$-23.17$&$-10.86$\\
007&$2.35$&$0.84$&$5.67$&$1.36$&$-15.47$&$-5.88 $\\
008&$5.77$&$2.46$&$2.71$&$1.49$&$-30.84$&$-14.45$\\
009&$2.87$&$0.86$&$5.57$&$1.65$&$-25.22$&$-7.26 $\\
\noalign{\smallskip}
\hline\hline
\noalign{\smallskip}
\multicolumn{7}{c}{Only WHIM}\\
\noalign{\smallskip}
Subvol.&$\max(y_{\rm max})$&$\min(y_{\rm max})$&$\max(|b|_{\rm max})$&$\min(|b|_{\rm max})$&$\min(\Delta T_{\rm min})$&$\max(\Delta T_{\rm min})$\\
\noalign{\smallskip}
&$[10^{-6}]$&$[10^{-6}]$&$[10^{-6}]$&$[10^{-6}]$&$[\mu K]$&$[\mu K]$\\
\hline
\noalign{\smallskip}
001&$1.94$&$0.80$&$3.26$&$1.57$&$-14.77$&$-5.99 $\\
002&$1.76$&$0.86$&$6.06$&$3.22$&$-25.99$&$-12.93$\\
003&$1.64$&$0.83$&$3.91$&$1.78$&$-16.20$&$-8.18 $\\
004&$1.69$&$0.88$&$7.69$&$3.54$&$-8.37 $&$-3.85 $\\
005&$1.55$&$0.67$&$2.70$&$1.38$&$-12.47$&$-7.05 $\\
006&$1.35$&$0.49$&$2.72$&$1.46$&$-11.14$&$-2.80 $\\
007&$1.14$&$0.69$&$3.79$&$1.08$&$-9.32 $&$-5.51 $\\
008&$1.37$&$0.84$&$2.52$&$1.41$&$-10.88$&$-6.74 $\\
009&$1.31$&$0.44$&$5.32$&$1.59$&$-18.85$&$-6.69 $\\
\noalign{\smallskip}
\hline\hline
\label{table:SZ_with_Planck}
\end{tabular*}
}
\normalsize
\rm
\end{center}
\end{table*}

\section{Discussion and Conclusions}
\label{section7}

Based on the nine supercluster regions extracted from the MNU simulation, our
main conclusion is that a Comptonization $y$-parameter as large as the one
measured in CrB-SC by MITO/VSA is probably produced by galaxy clusters ($M \ge
\pot{5}{13}$~$\munit$).
When excluding the tSZ component associated with galaxy clusters, the maximum
$y$-values are typically of the order of $y\lesssim\pot{2}{-6}$.
Exceptionally, we do obtain orientations in which the maximum value is
compatible at the lower end of the 1~sigma level (with a largest value
  of $y\sim\pot{2.84}{-6}$). However, these ``privileged'' orientations
  represent less than  0.3\% of all the cases.
When we take into account the actual posterior distribution
  from the observations as described in equation~\ref{prob}, the
  probability of finding \emph{CrB-like} features in the maps without
  clusters is $3.2\%$.
Moreover, those values are comparable to (and in some cases smaller
  than) the expected contribution to the los SZ integral by background clusters
  \cite[see e.g.][]{Springeletal01,Roncarellietal07}. Hence, it seems more
  plausible that the $y$-signal observed in the CrB-SC is caused by a cluster of
  galaxies that does not belong to the supercluster, and that is located further
  along the line of sight, instead of the hypothesis that the spot is being
  produced by small clusters or galaxy groups associated to the supercluster.

On the other hand, and if our simulations provide a suitable description
  to the gas physics, the WHIM phase, defined as gas with $10^5K<T<10^7K$,
  cannot build up a tSZ signal compatible with the observations. 
  If we take into account the posterior
distribution from the observations, the probability of finding tSZ
signals compatible with the observations (see eq.~\ref{prob}) is
smaller than $1\%$. Hence, although we can not rule out a contribution
of WHIM from the supercluster, it is unlikely the cause of the spot in
CrB.

Including the kSZ component does not modify this conclusion. However, the
simulations show that the kinematic component of the SZE plays an important role
in building up the total temperature decrement in the maps with clusters
removed. Whereas in clusters kSZ is about one order of magnitude lower than tSZ,
when clusters are excluded from the analysis, both SZ components are comparable.
This role is also noticeable in the temperature decrement probability
distribution, showing that in order to study the SZ effect due to WHIM, kSZ should
not be neglected. This idea seems to be supported by the work done by
\citet{Atrio08}, where it is shown that the kSZ effect originates in filaments
of gas of certain characteristics could yield temperature decrements comparable
to those of CrB. However, it should be noted that we have not been able to
reproduce the temperature anisotropies they predict, considering galaxy groups
plus WHIM in our simulations.

Our main conclusion might in principle depend on the cosmology. The
MNU simulation has been obtained using a rather large normalization
power spectrum to set up the initial conditions. Current WMAP5 results
indicate smaller values for both $\sigma_{8}$ and $\Omega_{\rm m}$
than the ones we used in the simulation. This will translate into less
evolved structures, where the abundance of massive clusters in the
same volume is considerably suppressed as was shown in \citet{Yepes07}
from a set of different MareNostrum simulations with WMAP1 and WMAP3
cosmological parameters. This means that, attending only to the
cosmology, the results shown in this paper constitute an upper limit
to the probability of finding features like the CrB spot for the same
cosmic volume sampled. We cannot neglect either the effect of cosmic
variance: using a larger volume we would have been able to extend the
sample of superclusters and thus improve the statistics in order to
obtain a better probability function. However, according to the
variance shown in the simulation, we do not expect significant changes
in the results.

The other caveat concerns the modelling of the physics of baryons in
clusters. MNU is a pure adiabatic gasdynamical simulation where radiative
processes are not taken into account. The inclusion of radiative cooling, UV
photoionization and star formation would change the properties of the gas inside
clusters and the corresponding SZ effect. How large this change is not a
simple answer since we are not yet able to perform large scale simulations of
the size of the MareNostrum Universe but including also cooling and star
formation effects.
According to \citet{CenOstriker06}, the inclusion of these feedback
effects does not significantly modify the fraction of baryons expected
in the WHIM.  However, feedback does slightly  modify the differential mass
fraction, heating the gas from the warm phase ($T\lesssim 10^{5}K$)
into the WHIM phase, while the hot phase ($T\gtrsim 10^{7}K$) remains
unchanged. Given that the greatest SZ effect from the WHIM originates in the
phase with $T\sim 10^{7}K$ \citep[e.g.][]{HMetal06}, our results seem
fairly robust against feedback. 
Nonetheless, it should be further studied, especially since shocks
could  play a very important role in heating the gas in the CrB
cluster and hence be the cause of the spot. Despite the lack of large
scale simulations with feedback effects, one possibility for this
study is to perform a re-simulation of a volume-limited sample of
clusters in which considerably more physical processes are to be
included. This will allow us to estimate the relative change in the
determination of the SZ effect with respect to the adiabatic runs and
estimate the effect of shocks. This is a future project.

All in all, and according to the description provided by the MNU 
gasdynamical simulations, we conclude that the observed thermal SZ 
component of the CrB H-spot most probably arises from one (or several) 
unknown galaxy cluster(s) along the line of sight, which in principle 
are not neccesarily associated with the CrB supercluster. In order to 
explore this idea further, we have started an observational programme 
to perform a detailed characterization of the properties of the galaxy 
populations along the line of sight of the CrB spot 
\citep[see][]{Carmen09}.

\section*{Acknowledgements}
This work has been partially funded by project AYA2007-68058-C03-01 
of the spanish Ministry of Science and Innovation.
This work has been supported by the Spanish-Italian "Accion Integrada" 
HI2004-0004.
Gustavo Yepes would like to thank MEC (Spain) for financial support
under project numbers FPA2006-01105 and AYA2006-15492-C03.
GL, MDP, LL and SDG have been supported by funding from Ateneo
2006-C26A0647AJ and Azioni Integrate Italia-Spagna IT2196.
The simulations used in this work are part of the MareNostrum
Numerical Cosmology Project that is under operation at the BSC-CNS
(Barcelona Supercomputing Center - Centro Nacional de
Supercomputaci\'on). Some of the data analyses of these simulations
were done also at NIC Julich and at LRZ (Germany).
We thank N.~Afshordi for useful suggestions about the kSZ component in
the early stages of this work.
JARM is a Ram\'on y Cajal fellow of the Spanish Ministry of Science and
Innovation.
We thank the anonymous referee for his/her comments, which helped to improve the
quality of this article.
We thank J.~Betancort-Rijo for his useful comments.
%


\label{lastpage}
\end{document}